\documentclass{LMCS}

\def\doi{8(3:18)2012}
\lmcsheading%
{\doi}
{1--27}
{}
{}
{Jul.~19, 2006}
{Sep.~19, 2012}
{}

\usepackage{enumerate}
\usepackage{hyperref}

\usepackage{amsmath}
\usepackage{amsfonts}
\usepackage{amssymb}
\usepackage{logic9-thm}

\newcommand{\shd}{\mathtt{sh\$}}
\newcommand{\shc}{\mathtt{shc}}

\renewcommand{\th}{\theta}

\newcommand{\Mg}{\mathcal{M}^g}
\newcommand{\Ml}{\mathcal{M}^l}

\newcommand{\RR}{\mathbb{R}}

\newcommand{\sExp}[2]{Exp(#1,#2)} 
\newcommand{\gamepar}[2]{{G_{#1,#2}}}
\newcommand{\game}{\gamepar{\Aa}{w}}
\newcommand{\adam}{Adam}
\newcommand{\ewa}{Eve}
\newcommand{\dUU}{\widetilde{\UU}}
\newcommand{\UU}{\mathbb{U}}
\newcommand{\hact}[1]{\stackrel{#1}{\hookrightarrow}}
\newcommand{\dact}[2]{\overset{#1}{\underset{#2}{\longrightarrow}}}
\newcommand{\mactS}{\stackrel{\Sigma^*}{\twoheadrightarrow}}
\newcommand{\fract}{\mathtt{fract}}
\newcommand{\reg}{\mathtt{reg}}
\newcommand{\bS}{\overline \S}
\newcommand{\delay}{\mathtt{delay}}

\newcommand{\dto}[1]{\act{(\mathtt{delay},#1)}}
\newcommand{\rplus}{{\mathbb R}_+}

\newcommand{\rreset}{\mathtt{reset}}
\newcommand{\nnop}{\mathtt{nop}}
\newcommand{\partto}{\stackrel{\cdot}{\to}}
\newcommand{\sqleq}{\sqsubseteq}
\newcommand{\actS}{\stackrel{\Sigma^*}{\twoheadrightarrow}}
\newcommand{\ddelay}{\mathtt{delay}}
\newcommand{\hCc}{\widehat{C}}
\newcommand{\hGg}{\widehat{G}}
\newcommand{\ccc}{\widehat{c}}
\newcommand{\up}{\!\!\uparrow}
\newcommand{\Exp}[2]{exp(#1,#2)} 

\begin{document}

\title{Weak Alternating Timed Automata} 
\author[P.~Parys]{Pawe\l{} Parys\rsuper a}
\address{{\lsuper a}Warsaw University, Poland}
\thanks{{\lsuper a}Author supported by Polish government grant
no. N206 008 32/0810.} 
\author[I.~Walukiewicz]{Igor Walukiewicz\rsuper b}
\address{{\lsuper b}CNRS and Bordeaux University, France}
\thanks{{\lsuper b}Author supported by project DOTS (ANR-06-SETI-003).}

\hypersetup{%
  pdftitle = {Weak Alternating Timed Automata},
  pdfkeywords = {verification, timed systems, alternating timed automata},
  pdfauthor = {Parys, Walukiewicz}
}


\begin{abstract}
  \noindent Alternating timed automata on infinite words are considered. The
  main result is a characterization of acceptance conditions for
  which the emptiness problem for these automata is decidable. This
  result implies new decidability results for fragments of timed
  temporal logics. It is also shown that, unlike for MITL, the
  characterisation remains the same even if no punctual constraints are
  allowed.
\end{abstract}

\maketitle

\section{Introduction}

Timed automata~\cite{AluDil94} are widely used models of real-time
systems. They are obtained from finite automata by adding clocks that can
be reset and whose values can be compared with constants.  The crucial
property of timed automata is that their emptiness is
decidable. Some other properties, like universality, are undecidable
though. Alternating timed automata have been introduced
in~\cite{LasWal05,OW05} following a sequence of
results~\cite{AJ98,AJ01,OW04} indicating that a restriction to one
clock can influence decidability. Indeed, the emptiness and
universality  problems for
one clock alternating timed automata are decidable over finite words. On the
contrary, over infinite words both problems remain undecidable even
for automata with one clock~\cite{OW07,LW08}. All undecidability
arguments rely on the ability to express ``infinitely often''
properties. Our main result shows that once these kind of properties
are forbidden the emptiness problem is decidable.

To say formally  what are ``infinitely often'' properties we look at
the theory of infinite sequences. We borrow from that theory the notion
of an index of a language. It is known that the index hierarchy is
infinite with ``infinitely often'' properties almost at its bottom.
From this point of view, the undecidability result mentioned above
leaves open the possibility that safety properties and ``almost
always'' properties can be decidable. This is indeed what we prove
here. 

The automata theoretic approach to temporal logics~\cite{VarWop84a} is by
now a standard way of understanding these formalisms. For example, we
know that the modal $\mu$-calculus corresponds to all automata, and
LTL to very weak alternating automata, or equivalently, to counter-free
nondeterministic automata~\cite{WilkeHab89}. By translating a logic to
automata we can clearly see combinatorial challenges posed by the
formalism. We can also abstract from irrelevant details, such as a
choice of operators for a logic. This approach was very beneficial for
the development of logical formalisms over sequences.

An automata approach has been missing in timed models for an obvious
reason: no standard model of timed automata is closed under boolean
operations. Event-clock automata~\cite{AluFixHen97} may be considered
as an exception, but the price to pay is a restriction on the use of
clocks.  Alternating timed automata seem to be a good model, although
the undecidability result over infinite words shows that the situation
is more difficult than for finite sequences. 

 The idea of restricting to one clock automata
dates back at least to~\cite{HJ96}. Alternating timed automata were
studied in a number of papers~\cite{LW08,OW07,AOQW07,ADOQW08}. Our
main result is that the emptiness problem for alternating timed
automata with one clock and ``almost always'' conditions is
decidable. A particular case of such automata is when all the states
are accepting.  This case was considered by Ouaknine and
Worrell~\cite{OWsafety06} who have shown decidability of the emptiness
problem under some additional restriction on the form of transitions.

The above mentioned result of Ouaknine and Worrell allowed them to
identify a decidable fragment of MTL called Safety MTL.  In the
present paper we show that our main theorem allows to get a decidable
fragment of TPTL~\cite{AH94} with one variable, that we call
Constrained TPTL. This fragment contains Safety MTL, allows all
``eventually'' formulas, and more liberal use of clock constraints. Its
syntax has also some similarities with another recently introduced
logic: FlatMTL~\cite{BMOW07,BMOW08}. We give some elements of
comparison between the logics later in the paper. In brief, the reason
why Constrained TPTL is not strictly more expressive than FlatMTL is
that the later includes MITL~\cite{alurfedhen96}. This is a sub-logic
of MTL where punctualilty constraints are not allowed.

The case of MITL makes it natural to ask what happens to alternating
timed automata when we disallow punctual constraints. This is an
interesting question also because all known undecidability proofs have
used punctual constraints in an essential way. Our second main result
(Theorem~\ref{thm:undecidable_ne}), says that the decidability
frontier does not change even if we only allow to test if the value of
a clock is bigger than $1$. Put it differently, it is not only
the lack of  punctual constraints, but also the very weak syntax of the
logic that makes MITL decidable.

We should also discuss the distinction between continuous and
pointwise semantics. In the latter, the additional restriction is that
formulas are evaluated only in positions when an action happens. So
the meaning of $F_{(x=1)}\a$ in the continuous semantics is that in
one time unit from now formula $\a$ holds, while in the pointwise
semantics we additionally require that there is an action one time
unit from now.  Pointwise semantics is less natural if one thinks of
encoding properties of monadic predicates over reals. Yet, it seems
sufficient for descriptions of behaviors of devices, like timed
automata, over time~\cite{OW08}. Here we consider the pointwise
semantics simply because the emptiness of alternating timed automata
in continuous semantics is undecidable even over finite
words. At present it seems
that an approach through compositional methods~\cite{HR04} is more
suitable to deal with continuous semantics.

 Our work inserts itself also into the line of
research using well-quasi-orders to solve decidability
questions. Particularly close are models of lossy counter machines and
their duals: machines with incremental errors. Ouaknine and Worrell have
shown the undecidability of the emptiness problem for ATA over
infinite words by reduction to the repeated reachability problem for
incremental machines with occurrence testing (ICMOT)~\cite{OW06}. While
paper~\cite{BMOSW08} gives a finer analysis of the complexity of
several problems for ICMOT, using well-quasi orders it is easy to show
that the existence of a computation satisfying ``almost always'' property
is decidable for ICMOT. Nevertheless this observation does not imply
decidability of the same problem for ATA, whose structure is more
complicated. It is worth to mention that checking the existence of a
run satisfying ``almost always'' property is in general more difficult
than checking reachability.  Recall for example that the former
problem is not decidable for lossy counter machines~\cite{Mayr03},
while reachability is decidable for this model.  

The depth of nesting of positive and negative conditions of type
``infinitely often'' is reflected in the concept of the index of an
automaton. Wagner~\cite{Wag77}, as early as in 1977, established the
strictness of the hierarchy of indices for deterministic automata on
infinite words. Weak conditions were first considered by Staiger and
Wagner~\cite{WS74}. There are several results testifying their
relevance. For example Mostowski~\cite{Mos91} has shown a direct
correspondence between the index of weak conditions and the
alternation depth of weak second-order quantifiers. For recent results
on weak conditions see~\cite{Murlak08} and references therein. 
\medskip

The next preliminary section is followed by a presentation of
our main decidability result (Theorem~\ref{thm:decidable}).
Section~\ref{sec:logic} introduces Constrained TPTL, gives a
translation of the logic into a decidable class of alternating timed
automata, and discusses relations with FlatMTL. The last section
presents the accompanying undecidability result
(Theorem~\ref{thm:undecidable_ne}).

\section{Preliminaries}

A \emph{timed word} over a finite alphabet $\S$ is a sequence
\begin{equation*}
  w=(a_1,t_1)(a_2,t_2)\dots
\end{equation*}
of pairs from $\S\times\rplus$. We require that the sequence
$\set{t_i}_{i=1,2,\dots}$ is strictly increasing and unbounded. If
$t_i$ describes the time when event $a_i$ has occurred then these
restrictions say that there cannot be two actions at the same time instance
and that there cannot be infinitely many actions in a finite time
interval (non Zeno behavior).

We will consider alternating timed automata (ATA) with one
clock~\cite{LW08}. Let $x$ be this clock and let $\F$ denote the set of
all comparisons of $x$ with constants, eg. $(x<1\land x\geq 0)$. 

A one-clock  ATA over an alphabet $\S$ is a tuple
\begin{equation*}
  \Aa=\struct{Q,\S,q_o,\delta,\W: Q\to\mathbb{N}},
\end{equation*}
where $Q$ is a finite set of states and $\W$ determines the parity
acceptance condition. The transition function of the automaton
$\delta$ is a finite partial function
\begin{equation*}
  \delta:Q \times \Sigma \times \F \partto \Bb^+(Q \times 
  \set{\nnop,\rreset}),
\end{equation*}
where $\Bb^+(Q \times \set{\nnop,\rreset})$ is the set of positive
boolean formulas over atomic propositions of the form $\top$,
$\bot$, and $(q,f)$ with $q\in Q$ and $f\in
\set{\nnop,\rreset}$.

Intuitively, automaton being in a state $q$, reading a letter $a$, and
having a clock valuation satisfying $\th$ can proceed according to the
positive boolean formula $\d(q,a,\th)$. It means that if a formula is
a disjunction then it chooses one of the disjuncts to follow, if it is
a conjunction then it makes two copies of itself each following one
conjunct. If a formula is ``atomic'', i.e., of the form $(q,\rreset)$
or $(q,\nnop)$ then the automaton changes the state to $q$ and either
sets the value of the clock to $0$ or leaves it unchanged,
respectively. To simplify the definition of acceptance there is also
one more restriction on the transition function:
\begin{quote}
  \emph{(Partition)} For every $q\in Q$, $a\in \S$ and $v\in\rplus$,
  there is at most one $\th$ s.t. $\d(q,a,\th)$ is defined, and $v$
  satisfies $\th$.
\end{quote}
It is easy to transform an automaton to this form. 

The \emph{acceptance condition} of the automaton determines which infinite
sequences of states (runs of the automaton) are accepting.  A sequence
$q_1,q_2,\dots$ satisfies:
\begin{iteMize}{$\bullet$}\label{weak-parity}
\item \emph{weak parity condition} if $\min\set{\W(q_i) :i=1,2,\dots}$ is
  even,

\item \emph{strong parity condition} if $\liminf_{i=1,2,\dots} \W(q_i)$ is
even.
\end{iteMize}
Observe that the difference between weak and strong conditions is that
in the weak case we consider all occurrences of states and in the
strong case only those that occur infinitely often. In this paper we
will mostly consider automata with weak conditions. Whenever we will be
considering strong conditions we will say it explicitly.

For an alternating timed automaton $\Aa$ and a timed word $
w=(a_1,t_1)(a_2,t_2)\dots$ we define the \emph{acceptance game
  $\game$} between two players: \adam\ and \ewa.  Intuitively, the
objective of \ewa\ is to accept $w$, while the aim of \adam\ is the
opposite.  A \emph{play} starts at the initial configuration $(q_0,0)$.  It
consists of potentially infinitely many phases.  The $(k{+}1)$-th
phase starts in $(q_k,v_k)$, ends in some configuration
$(q_{k{+}1},v_{k{+}1})$ and proceeds as follows.  Let $v' := v +
t_{k{+}1}-t_k$.  Let $\th$ be a unique (by the partition condition)
constraint such that $v'$ satisfies $\th$ and $\d(q_k,a_{k+1},\th)$
is defined; if there is no such $\th$ then \ewa{} is blocked.  Now the
outcome of the phase is determined by
the formula $b=\d(q_k,a_{k+1},\th)$. There are four cases:
\begin{iteMize}{$\bullet$}
\item $b = b_1 \land b_2$: \adam\ chooses one of subformulas $b_1$,
$b_2$ and the
play continues with $b$ replaced by the chosen subformula;
\item
$b = b_1 \lor b_2$: dually, \ewa\ chooses one of subformulas;
\item
$b = (q, f) \in Q \times \set{\nnop,\rreset}$: the phase ends
with the result
$(q_{k{+}1},v_{k{+}1}) := (q,f(v'))$ and
a new phase starts from this configuration;
\item $b=\top,\bot$: the play ends.
\end{iteMize}
The winner of such a play is \ewa\ if she is not blocked, and the
sequence ends in $\top$, or it is infinite and the states appearing in
the sequence satisfy the acceptance condition of the automaton.

Formally, a play is a finite sequence of consecutive game positions
of the form $\langle k, q, v \rangle$ or $\langle k, q, v, b \rangle$,
where $k$ is the phase number, $b$ a boolean formula, $q$ a location
and $v$ a valuation.  A \emph{strategy} of Eve is a mapping which
assigns to each such sequence ending in Eve's position a next move of
Eve.
A strategy is \emph{winning} if all the plays respecting the strategy are
winning.

\begin{defi}[Acceptance]
  An automaton $\Aa$ {\em accepts} $w$ iff \ewa\ has a winning
  strategy in the game $\game$.  By $L(\Aa)$ we denote the language of
  all timed words $w$ accepted by $\Aa$.
\end{defi}

The \emph{Mostowski index} of an automaton with the, strong or weak,
acceptance condition given by $\W$ is the pair consisting of the
minimal and the maximal value of $\W$: $(\min(\W(Q)),\max(\W(Q)))$. We
may assume without a loss of generality that $\min(\W(Q)) \in \set{0,
  1}$. (Otherwise we can scale down the rank by $\W(q ) := \W(q)
-2$.)  Automata with strong conditions of index $(0,1)$ are
traditionally called B\"uchi automata and their acceptance condition
is given by a set of accepting states $Q_+\incl Q$; in our
presentation these are states with rank $0$.



\section{Decidability for one-clock timed automata}

We are interested in the emptiness problem for one clock ATA. As it
was mentioned in the introduction, the problem is undecidable for
automata with strong B\"uchi conditions (strong $(0,1)$
conditions). Here we will show a decidability result for automata with
weak acceptance conditions of index $(0,1)$.

\begin{thm}\label{thm:decidable}
  It is decidable whether a given one-clock alternating timed
  automaton with weak $(0,1)$ condition accepts some non Zeno timed
  word. The complexity of the problem is non-primitive recursive.
\end{thm}
The lower bound for the complexity holds already for automata over
finite words~\cite{LW08}. So in the rest of this section we give a
decidability proof. 

Before we start, it will be useful to make a couple of remarks that
allow to restrict the form of automata. A weak $(0,1)$ automaton can
be also presented as an automaton with a strong $(0,1)$ condition
where all transitions from an accepting state, state of rank $0$, go
only to accepting states.  Indeed, once the automaton sees a state of
priority $0$ then any infinite run is accepting (but there may be runs
that get blocked). In the following we will write $Q_+$ for accepting
states and $Q_-$ for the other states. For automata presented in this
way the strong $(0,1)$ condition says simply: there are
only finitely many states from $Q_-$ in the run. So the automaton
accepts if Eve has a strategy to reach $\top$, or to satisfy this
condition.

We can also make some restrictions on a form of the transition
function.  We can require that every boolean formula that appears as a
value of the function is in a disjunctive normal form.
Moreover, we can eliminate the $\bot$ and $\top$ propositions. Proposition
$\bot$ can be simulated by a state $q_\bot$ from which there is no transition,
and $\top$ by an accepting state $q_\top$ on which the automaton loops on all letters.
Observe that this is fine as
we have put no restriction on transitions going to accepting states. 
Finally, we can assume that every disjunct of every transition of
$\Aa$ has some pair with $\rreset$ and some pair with
$\nnop$. This can be guaranteed by adding conjuncts
$(q_\top,\nnop)$ and $(q_\top, \rreset)$.


To fix  the notation we take a one clock ATA in a form as described above:
\begin{equation*}
    \Aa=\struct{Q,\S,q_o,\delta,Q_+\incl Q}.
\end{equation*}
This means that for every $q$, $a$, and $\theta$, the formula
$\d(q,a,\theta)$ is in a disjunctive normal form; every disjunct
contains a pair with $\nnop$ and a pair with $\rreset$; there are no
$\top$ or $\bot$; if $q\in Q_+$ then only states from $Q_+$ appear in
the formula;

Our first step will be to construct some infinite transition system
$\Hh(\Aa)$, so that the existence of an accepting run of $\Aa$ is
equivalent to the existence of some good path in $\Hh(\Aa)$. In the second
step we will use some structural properties of this transition system
to show decidability of the problem stated in the theorem.

\subsection{An abstract transition system}

The goal of this subsection is to define a transition system $\Hh(\Aa)$
such that existence of an accepting computation of $\Aa$ is reduced to
existence of some special infinite path in $\Hh(\Aa)$
(Corollary~\ref{cor:autom_to_H}). This system will be some abstraction
of the transition system of configurations of $\Aa$. While $\Hh(\Aa)$
will be infinite, it will have some well-order structure and other
additional properties that will permit to analyze it.

First, consider an auxiliary labeled transition system $\Ss(\Aa)$
whose states are finite sets of configurations, i.e., finite sets of
pairs $(q,v)$, where $q \in Q$ and $v \in \rplus$.  The
initial position in $\Ss(\Aa)$ is $P_0 = \{ (q_0,0) \}$ and
there are transitions of two types $P\hact{t}P'$ and
$P\hact{a}P'$.  Transition  $P\hact{t}P'$ is in $\Ss(\Aa)$  iff
$P'$ can be obtained from $P$ by changing every configuration
$(q,v)\in P$ to $(q,v+t)$. Transition $P\hact{a}P'$ is in $\Ss(\Aa)$ iff 
$P'$ can be obtained from $P$ by the following nondeterministic process:
\begin{iteMize}{$\bullet$}
\item
First, for each $(q,v) \in P$, do the following:
\begin{iteMize}{$-$}
\item
let $b=\d(q,a,\theta)$ for the uniquely determined $\theta$ satisfied in
$v$,
\item
choose one of disjuncts of $b$, say
\[
(q_1,r_1) \land \ \cdots \ \land (q_k,r_k)
\ \ \ \ \ (k > 0),
\]
\item
let $\mathrm{Next}(q,v) = \{ (q_i,r_i(v)) : i = 1 \ldots k \}$.
\end{iteMize}
\item
Then, let $P' := \bigcup_{(q,v) \in P} \mathrm{Next}(q,v)$.
\end{iteMize}
Observe that there may be no $P'$ such that $P\hact{a} P'$ because for
some $(q,v)\in P$ the value $\d(q,a,\theta)$ required above is not defined.

\begin{defi}
  We will call a sequence $P_0,P_1,\dots$ of the states of $\Ss(\Aa)$
  \emph{accepting} if the states from $Q_-$ appear only in a finite
  number of $P_i$.
\end{defi}

\begin{lem}\label{lemma:acceptance}
  $\Aa$ accepts an infinite timed word $(a_0,t_0)(a_1,t_1)\dots$ iff
  there is an accepting sequence in $\Ss(\Aa)$:
  \begin{equation*}
    P_0\hact{t_0} P_1\hact{a_0} P_2 \hact{t_1} P_3\hact{a_1} P_4\dots
  \end{equation*}
\end{lem}
\proof
  The right to left implication is obvious. For the left to right
  implication, recall that acceptance of a word by an automaton is
  defined as existence of a winning strategy for \ewa{} in the
  acceptance game.  This is a game with B\"uchi conditions, so if
  \ewa{} has a winning strategy, then she has a memoryless winning
  strategy. This strategy gives a run of the form required by the
  lemma.
\qed

Our next goal is to remove time labels on transitions. But we cannot
just erase them, as then we will not be able to say if a word is
Zeno or not. We start by introducing regions.

Let $d_{max}$ denote the biggest constant appearing in $\delta$,
i.e., the transition function of the automaton. Let set $\mathtt{reg}$
of \emph{regions} be a partition of $\mathbb{R}_+$ into
$2\cdot(d_{max}+1)$ sets as follows:
\begin{equation*}
\mathtt{reg}:=\{\{0\},(0,1),\{1\},(1,2),\dots,
(d_{max}-1,d_{max}),\{d_{max}\},(d_{max},+\infty)\}.
\end{equation*}
There are three kinds of regions: bounded intervals (denoted
$\mathtt{reg}_I$), one-point regions (denoted $\mathtt{reg}_P$), and
one unbounded interval $(d_{max},+\infty)$.  We will use the notation
$\Ii_{i}$ for the region $(i-1,i)$. In a similar way, $\Ii_{\infty}$
will stand for $(d_{max},+\infty)$. For $v \in \rplus$, let $\reg(v)$
denote the region $v$ belongs to; and let $\fract(v)$ denote the
fractional part of $v$.

Let us try to give an intuition behind the way time information will
be eliminated. Recall that a state $P$ is a finite set of pairs
$(q,v)$. If $v\in I_\infty$ then the precise value of $v$ does not
matter from the point of view of the automaton. For other values it is
important to look at their fractional parts. Among all $v\not\in
\Ii_\infty$ appearing in $P$ take the one with the biggest fractional
part. Then, by making the time pass we can get $v$ to a new region
without changing the regions of valuations with smaller, but positive,
fractional parts. Intuitively this is the smallest delay that makes a
visible change to $P$. We will introduce a special label to signal
when time progresses in this way. As integer valuation would force us
to introduce a cumbersome case distinction we will set things so that
they can be avoided.

These remarks lead us to consider a new alphabet:
\begin{equation*}
  \bS=\S\cup\set{(\delay,\e)}\cup(\set{\ddelay}\times\S),
\end{equation*}
and three new kinds of transitions.

Transition on $a$ will do the action and make some time pass without
any valuation changing the region.
\begin{quote}
  $P\act{a} P'$ if $P\hact{a}P_1\hact{t_1}P'$ for some
  $P_1,$ and $t_1>0$ such that for every $(q,v)\in P$, the value
  $v+t_1$ is in the same region as $v$.
\end{quote}
 
For a transition on a letter $(\ddelay,\e)$, pick a valuation $v$
among these with $\reg(v)\not=\Ii_{\infty}$ with a maximal
$\fract(v)$.  The transition will make the time pass so that $v$ goes
to the next interval region but all valuations with smaller fractional
parts do not change their regions:
\begin{quote}
  $P \dto{\e} P'$ if $P\hact{t_1}P'_1\hact{t_2}P'$ for some $P'_1$ and
  $t_1,t_2>0$ such that there is $(q,v)\in P$, with $v+t_1$ being an
  integer and $v+t_1+t_2$ in the following interval region. Moreover,
  for all $(q',v')\in P$ if $\fract(v)\not=\fract(v')$ then the value
  $v'+t_1+t_2$ is in the same region as $v'$.
\end{quote}

Finally, we come to the most complex $(\ddelay,a)$ transition. Even
though we did not allow transitions $(\ddelay,\e)$ to reach one-point
regions, it is still  important to be able to execute actions in those
regions. A transition on $(\ddelay,a)$ permits to reach a one-point
region, execute the action, and leave the region.

\begin{quote}
  $P \dto{a} P'$ if $P\hact{t_1}P_1\hact{a}P_2\hact{t_2}P'$ for some
  $P_1,P_2$ and $t_1,t_2>0$ such that there is $(q,v)\in P$, with
  $v+t_1$ being an integer and $v+t_1+t_2$ in the following interval
  region. Moreover for all $(q',v')\in P$ if
  $\fract(v)\not=\fract(v')$ then the value $v'+t_1+t_2$ is in the
  same region as $v'$.
\end{quote}

The following lemma shows that with a new alphabet we can replace non
Zeno condition by a simple infinitary condition.

\begin{lem}\label{lem:autom_to_T}
  There is a non Zeno accepting sequence in $\Ss(\Aa)$:
  \begin{equation*}
    P_0\hact{t_0} P_1\hact{a_0} P_2 \hact{t_1} P_3\hact{a_1} P_4\dots
  \end{equation*}
  iff there is an accepting sequence
    \begin{equation*}
P_0\act{\s_0} P'_1\act{\s_1} P'_2\dots,
  \end{equation*}
  where $\s_0,\s_1,\dots\in \bS$ and $(\ddelay,\cdot)$ letters appear
  infinitely often in the sequence.\qed
\end{lem}

The next step in the construction is to abstract from valuations in
the states of the transition system. Intuitively, we will replace every
valuation by its region. To compensate for erasing fractional parts,
we will also keep information about the relative order between them.
With the construction described in the definition below the states
become words from
\begin{gather*}
  \Lambda_I^*\cdot \Lambda_\infty,\\
  \text{where $\Lambda_I=\Pp(Q\times\reg_I)$ and
$\Lambda_\infty=\Pp(Q\times\set{\infty})$.}
\end{gather*}

\begin{defi}
\label{d:H}
For a state $P$ of $\Ss(\Aa)$ we define a word $H(P)$ from
$\Lambda_I^*\cdot\Lambda_\infty$ as
the one obtained by the following procedure:
\begin{iteMize}{$\bullet$}
\item
replace each $(q,v) \in P$ by a triple
$\langle q, \reg(v), \fract(v) \rangle$ if $v\leq d_{max}$ 
(this yields a finite set of triples)
\item
sort all these triples w.r.t.\ $\fract(v)$
(this yields a finite sequence of triples)
\item
group together triples having the same value of $\fract(v)$
(this yields a finite sequence of finite sets of triples)
\item
forget $\fract(v)$, that is, change every triple
$\langle q, \reg(v), \fract(v) \rangle$ into a
pair $(q,\reg(v))$
(this yields a finite sequence of finite sets of pairs,
a word in $\Lambda_I^*$).
\item Add at the end the letter $(\set{q : (q,v)\in P, v>
    d_{max}},\Ii_\infty)\in\Lambda_\infty$. 
\end{iteMize}
\end{defi}

Finally, we can define $\Hh(\Aa)$. 

\begin{defi}
$\Hh(\Aa)$ is a transition system which has $\Lambda^*_I\times\Lambda_{\infty}$ as a set of configurations, and for every letter
$\s\in\bar\S$ there is a transition $c\act{\s} c'$ if there are states
$P,P'$ of $\Ss(\Aa)$ such that $P\act{a} P'$ and $H(P)=c$,
$H(P')=c'$. 
\end{defi}

Direct examination of the definition gives us the following.

\begin{lem}\label{lemma:simulation}
  If $H(P_1)=H(P_2)$ and $P_1\act{\s}P'_1$ then $P_2\act{\s} P'_2$ with
  $H(P'_1)=H(P'_2)$.\qed 
\end{lem}

\begin{defi}
  We say that a path (equivalently: run, computation) in $\Hh(\Aa)$ is \emph{good}, if it passes
  through infinitely many transitions labeled by letters $(\delay,\cdot)$.
  We say that a path (equivalently: run, computation) in $\Hh(\Aa)$ is \emph{accepting}, if it is good
  and passes through only finitely many configurations containing states from $Q_-$.
\end{defi}

\begin{cor}\label{cor:autom_to_H}
  $\Aa$ accepts an infinite non Zeno timed word iff there is an accepting
  path in $\Hh(\Aa)$ starting in configuration
  $(\{(q_0,\Ii_1)\},\set{\es,\I_\infty})$.
\end{cor}
\proof
  $\Aa$ accepts a non Zeno word iff there is a path in $\Ss(\Aa)$
  satisfying the acceptance condition. By Lemma~\ref{lem:autom_to_T}
  it is equivalent to having a good path in $\Ss(\Aa)$ with
  transitions from the alphabet $\bS$ satisfying the acceptance
  conditions. Lemma~\ref{lemma:simulation} implies that this is
  equivalent to having an accepting path in $\Hh(\Aa)$.
\qed

We finish the section with a more explicit characterization of
transitions in $\Hh(\Aa)$ that will be used extensively in the
decidability proof.  The characterisation is spelled out in the next
three lemmas whose proofs are obtained directly from the definitions.

\begin{lem}\label{lem:tranz_dla_delay}
  Consider a state $(\l_1\dots\l_k,\lambda_\infty)$ of $\Hh(\Aa)$. If
  $k=0$ then there is no $(\ddelay,\cdot)$ transition from this
  state. Otherwise let $\l'_k=\set{(q,\Ii_{d+1}) : d<d_{max}, (q,\Ii_d)\in\l_k}$ and
  $\l'_{\infty}=\l_\infty\cup\set{(q,\Ii_\infty):
    (q,\Ii_{d_{max}})\in\l_k}$. In $\Hh(\Aa)$ there is exactly one
  transition on $(\delay,\epsilon)$:
\begin{eqnarray*}
   &&(\l_1\dots\l_k,\lambda_\infty)\act{(\mathtt{delay},\epsilon)} (\l'_k\l_1\dots\l_{k-1},\l'_{\infty})
	\qquad\textrm{if }\l'_k\not=\emptyset,\\
   &&(\l_1\dots\l_k,\lambda_\infty)\act{(\mathtt{delay},\epsilon)} (\l_1\dots\l_{k-1},\l'_{\infty})
	\qquad\textrm{otherwise}\rlap{\hbox to86 pt{\hfill\qEd}}.
\end{eqnarray*}
\end{lem}

In order to describe transitions of $\Hh(\Aa)$ on an action $a$, we
define an auxiliary notion of a transition from
$\l\in\Pp(Q\times\reg)$. By the partition condition, for every
$(q,r)\in\l$ there is at most one constraint $\th$ such that every
valuation in $r$ satisfies this constraint and $\delta(q,a,\th)$ is
defined. We choose a conjunct from $\delta(q,a,\th)$:
    $$(q_1,\nnop)\wedge\dots\wedge(q_l,\nnop)\wedge
      (q'_1,\mathtt{reset})\wedge\dots\wedge(q'_m,\mathtt{reset}).$$
From this choice we can obtain two sets:
$\mathrm{Next}(q,r)=\{(q_1,r),\dots,(q_l,r)\}$ and
$\mathrm{Next}_0(q,r)=\{(q'_1,\Ii_1),\dots,(q'_m,\Ii_1)\}$.
We put
\begin{gather*}
\l\act{a}_{\Aa}(\l',\gamma'),\quad\text{where}\\
   \lambda'=\bigcup_{(q,r)\in \lambda}  Next(q,r)\quad\text{and}\quad
   \gamma'=\bigcup_{(q,r)\in \lambda}   Next_0(q,r).
\end{gather*}
Observe that there are as many transitions $\act{a}$
from $\l$ as there are choices of different conjuncts for each pair
$(q,r)$ in $\l$. In particular there is no transition if for some pair
the transition function of the automaton is not defined.
Notice also that the clock after reseting, described by elements of $\gamma'$, is in interval $\I_1$, not in $\{0\}$;
this is because we describe a transition of the original automaton followed by a small time elapse.

\begin{lem}\label{lem:tranz_dla_lit}
  In  $\Hh(\Aa)$ transitions on an action $a$ have the form
 \begin{equation*}
   (\l_1\dots\l_k,\lambda_\infty)\act{a} (\gamma'\l'_1\dots\l'_k,\l'_{\infty}),
 \end{equation*}
 where $\l_i\act{a}_\Aa(\l'_i,\gamma'_i)$ and $\gamma'=\bigcup
 \gamma'_i$ (for $i=1,\dots,k,\infty$).\qed
\end{lem}
Note that neither $\gamma'$ nor any of $\l'_i$ may be empty. 

Finally, we have the most complicated case of $(\ddelay,a)$ action.
\begin{lem}\label{lem:tranz_dla_delay_lit}
  In $\Hh(\Aa)$ the transitions on an action $(\ddelay,a)$ have the form
  \begin{equation*}
    (\l_1\dots\l_k,\lambda_\infty)\dto{a}
    (\gamma'\l'_1\dots\l'_{k-1},\l''_{\infty}),
  \end{equation*}
  where the elements on the right are obtained by preforming the
  following steps:
\begin{iteMize}{$\bullet$}
\item First, we change regions in $\l_k$. Every pair $(q,\Ii_d)\in\l_k$
  becomes $(q,\{d\})$. Let us
  denote the result by $\l_k^1$.
\item For $i=1,\dots,k,\infty$ we take $\l'_i,\gamma'_i$ such that:
$\l^1_k\act{a}_{\Aa}(\l'_k,\gamma'_k)$ and  $\l_i\act{a}_{\Aa} 
  (\l'_i,\gamma'_i)$ for $i\not=k$.
\item We again increase regions in $\l'_k$: from $\set{d}$ they become
  $\Ii_{d+1}$, or $\Ii_\infty$ if $d=d_{max}$. 
\item We put $\gamma'=\bigcup\gamma'_i\cup\set{(q,\Ii_d) :
    (q,\{d\})\in \l'_k, d<d_{max}}$ and
  $\l''_{\infty}=\l'_\infty\cup\set{(q,\Ii_{\infty}):
    (q,\{d_{max}\})\in \l'_k}$.\qed
\end{iteMize}
\end{lem}

\noindent We write $c\to c'$,
$c\act{(\mathtt{delay},\cdot)}c'$,
$c\twoheadrightarrow c'$,
$c\stackrel{\Sigma^*}{\twoheadrightarrow} c'$ to denote that 
we may go from a configuration $c$ to
$c'$ using one transition, one transition reading a
letter of the form $(\mathtt{delay},\cdot)$, any number of transitions or any
number of transitions reading only letters from $\Sigma$, respectively.

\subsection{Finding an accepting path in $\Hh(\Aa)$.}

Here we overview the decision procedure, which is described in details in the next subsections. 
By Corollary~\ref{cor:autom_to_H}, our problem reduces to deciding
if in $\Hh(\Aa)$ there is a good path with only finitely
many appearances of states from $Q_-$. The decision procedure works in
two steps. In the first step we compute the set $\hGg$ of all
configurations of $\Hh(\Aa)$ from which there exists a good path. Observe
that if a configuration from $\hGg$ has only states from $Q_+$ then
there exists an accepting run from this configuration. So, in the second step it remains to
consider configurations that have states from both $Q_-$ and
$Q_+$. This is relatively easy as an accepting run from such a
configuration consists of a finite prefix ending in a configuration
without states from $Q_-$ and a good run from that
configuration. Hence, there is an accepting run from a
configuration iff it is possible to reach from it a configuration from
$\hGg$ that has only $Q_+$ states. Once we know $\hGg$, the later
problem can be solved using the standard reachability tree technique.

\subsection{Computing accepting configurations}

We start with the second step of our procedure as it is much easier
than the first one. We need to decide if from an initial state one can
reach a configuration from $\hGg$ having only $Q_+$ states.  We can
assume that we are given $\hGg$ but we need to discuss a little how it
is represented. It turns out that there are useful well-quasi-orders
on configurations that allow to represent $\hGg$ in a finitary
way~(Corollary~\ref{cor:zmniejszanie})

A \emph{well-quasi-order} is a relation with a property that for every
infinite sequence $c_1,c_2,\dots$ there exist indexes $i<j$ such that
the pair $(c_i,c_j)$ is in the relation.

The order we need is the relation, denoted $\preceq$, over
configurations of $\mathcal{H}(\mathcal{A})$: we put
$(\lambda_1\dots\lambda_k,\lambda_\infty)\preceq(\lambda'_1\dots\lambda'_{k'},\lambda'_\infty)$
if $\lambda_\infty\subseteq\lambda'_\infty$ and there exists a
strictly increasing function $f\colon\{1,\dots,k\}\to\{1,\dots,k'\}$
such that $\lambda_i\incl\lambda'_{f(i)}$ for each $i$. Observe that
here we use the fact that each $\l_i$ is a set so we can compare them
by inclusion.  This relation is somehow similar to the relation of
being a subsequence, but we do not require that the corresponding
letters are equal, only that the one from the smaller word is included
in the one from the greater word.  The first property of this order is
proved by a standard application of Higman's lemma.

\begin{lem}
  The relation $\preceq$ is a well-quasi-order.\qed
\end{lem}

The following shows an important interplay between $\preceq$ relation
and transitions of $\Hh(\Aa)$. 


\begin{lem}\label{lem:zmniejszanie}
  Let $c_1,c_1',c_2$ be configurations of $\mathcal{H}(\mathcal{A})$
  such that $c'_1\preceq c_1$. Whenever $c_1\twoheadrightarrow c_2$,
  then there exist $c'_2\preceq c_2$ such that $c'_1\twoheadrightarrow
  c'_2$ and the second computation has the length not greater than the
  first one. Similarly, when from $c_1$ there exists a good
  computation, then from $c'_1$ such a computation exists.
\end{lem}
\proof
  For the first statement of the lemma we will simulate one transition
  from $c_1$ by at most one transition from $c'_1$.  If $c_1\act{a}
  c_2$ then directly from Lemma \ref{lem:tranz_dla_lit} it follows
  that there is $c'_2\preceq c_2$ such that $c_2\act{a}c'_2$. When
  $c_1\dto{\e} c_2$ we have two cases depending on the relation
  between one before last element of the two configurations. To be
  more precise, suppose that
  $c_1=(\lambda_1\dots\lambda_k,\lambda_\infty)$ and
  $c'_1=(\lambda'_1\dots\lambda'_{k'},\lambda'_\infty)$. If
  $\lambda'_{k'}\subseteq\lambda_k$ then we may do
  $(\mathtt{delay},\epsilon)$ from $c'_1$ and we get $c'_2\preceq
  c_2$.  Otherwise already $c'_1\preceq c_2$, we do not do any action
  and take $c'_2=c'_1$. Similarly for $(\mathtt{delay},a)$: either we
  match it with $(\mathtt{delay},a)$ or just with $a$. An obvious
  induction gives a proof of the first statement.

  For the second statement we need to show that the computation from
  $c'_1$ obtained by matching steps as described above is good (if the
  one from $c_1$ has been good). This is not immediate as we remove
  some $(\delay,\cdot)$ letters in the matching computation. 

  Fix a good computation from $c_1$. Let $c_2$ be a configuration in a
  computation starting from $c_1$, and let $c'_2$ be the corresponding
  configuration in the matching computation from $c'_1$. To arrive at
  a contradiction assume that there are no delays after $c'_2$.  Let
  us denote $c_2'=(\lambda_1'\dots\lambda_{k'}',\lambda'_\infty)$ and
  $c_2=(\lambda_1\dots\lambda_k,\lambda_\infty)$. Because $c_2'\preceq
  c_2$, we know that $\lambda_{k'}'$ is covered by some $\l_i$, i.e.,
  $\l'_{k'}\subseteq\lambda_i$. Let us take the biggest possible $i$.
  If some $a$-action is done from $c_2$ then it is matched by
  an $a$-action from $c'_2$, and for the resulting configurations the
  inclusion is preserved. This can happen only finitely many times
  though, as there are infinitely many $(\delay,\cdot)$ actions after
  $c_2$.  If a $(\delay,\cdot)$ action is done from $c_2$ and $i=k$ then
  it is matched by a $(\delay,\cdot)$ action from $c'_2$, a
  contradiction with the choice of $c_i$. If $i<k$ then the element
  $\l'_{k'}$ is left on its position in $c'_2$, while in $c_2$ we
  remove $\l_k$, hence $\l_i$ covering $\l'_{k'}$ gets closer to the
  end of the sequence. Repeating this argument, we get that the
  covering $\l_i$ finally becomes the last element and the previous
  case applies.
\qed

\begin{cor}\label{cor:zmniejszanie}
  The set $\hGg$ is downward closed, so it can be described by the
  finite set of minimal elements that do not belong to it.\qed
\end{cor}

As we have mentioned before, there is a good accepting computation from a
configuration iff it is possible to reach from it a configuration from
$\hGg$ that has only $Q_+$ states. The following lemma says that this
property is decidable.

\begin{lem}\label{lemma:reach-tree}
  Let $X$ be a downward closed set of configurations of $\Hh(\Aa)$, represented by the (finite) set of all its minimal elements. 
  It is decidable
  whether from a given configuration one can reach a configuration in a given set $X$ that
  has only states from $Q_+$.
\end{lem}

\proof
  We will use a standard reachability tree argument. The reachability
  tree is a tree in which the initial configuration is in the root,
  and every configuration has as children all configurations that
  may be reached by reading one letter. The algorithm constructs a
  portion $t$ of the tree according to the following rule: do not add
  a node $c'$ to $t$ in a situation when among its ancestors there is
  some $c\preceq c'$. Each path of $t$ is finite because $\preceq$ is
  a well-quasi-order. Furthermore, since the degree of every node is
  finite, $t$ is a finite tree. Then we check $t$ for a
  configuration from $X$ without states from $Q_-$.

  We only need to prove that, if in the whole reachability tree there
  is a configuration as above (which means that
  $\mathcal{H}(\mathcal{A})$ may accept), then there is also some in $t$. Let
  $c$ be such a configuration reachable from the initial configuration of
  $\mathcal{H}(\mathcal{A})$ by a path $\pi$ of the shortest
  length. Assume that $c$ is not in $t$, i.e. there are two nodes on
  $\pi$, say $c_1$ and $c_2$, such that $c_1$ is an ancestor of $c_2$
  and $c_1\preceq c_2$ (i.e. $c_2$ was not added to $t$). Then from
  Lemma \ref{lem:zmniejszanie}, there exists $c'\preceq c$ that may be
  reached from $c_1$ and the path from $c_1$ to $c'$ will be not longer
  than that from $c_2$ to $c$. So the path leading to $c'$ from
  the initial configuration is strictly shorter than $\pi$. Moreover,
  as $c'\preceq c$ and $X$ is downward closed, we immediately deduce
  that $c'\in X$, and $c'$ does not contain states from $Q_-$ which is
  a contradiction.
\qed

\subsection{Computing $\hGg$}

In this subsection we deal with the main technical problem of the
proof that is computing the set $\hGg$ of all configurations from
which there exist a good computation. We will actually compute the
complement of $\hGg$. While we will use well-orderings in the proof,
standard termination arguments do not work in this case. We will need
to examine more closely the definition of $\Hh(\Aa)$ an in particular
the mechanics of its transition as described in
Lemmas~\ref{lem:tranz_dla_delay},~\ref{lem:tranz_dla_lit}, and~\ref{lem:tranz_dla_delay_lit}.

We write $X\up$ for an upward closure of set $X$,
\begin{equation*}
X\up=\{c:\exists_{c'\in X} c'\preceq c\}.
\end{equation*}
Observe that by Corollary~\ref{cor:zmniejszanie} the complement of $\hGg$
is upward closed.

Let set $pre_\mathtt{delay}^\forall$ (respectively
$pre_{\Sigma^*}^\forall$) contain all configurations, from which after
reading any letter $(\mathtt{delay},\cdot)$ (any number of letters from
$\Sigma$), we have to reach a configuration from $X$,
\begin{eqnarray*}
&&pre^\forall_\mathtt{delay}(X)=
  \{c:\forall_{c'}(c\act{(\mathtt{delay},\cdot)}c'\Rightarrow c'\in X)\},
\\
&&pre^\forall_{\Sigma^*}(X)=
  \{c:\forall_{c'}(c\stackrel{\Sigma^*}{\twoheadrightarrow}c'\Rightarrow c'\in X)\}.
\end{eqnarray*}
Now we can use these $pre$ operations to compute a sequence of sets of
configurations
\begin{equation*}
Z_{-1}=\emptyset, \qquad
Z_i=pre_{\Sigma^*}^\forall(pre^\forall_\mathtt{delay}(Z_{i-1}\up)).
\end{equation*}

It is important that we may effectively represent and compare all the
sets $Z_i\up$. Because the relation $\preceq$ is a well-quasi-order,
any upward closed set $X\up$ may be represented by finitely many
elements $c_1,\dots,c_k$ (called \emph{generators}) such that
$X\up=\{c_1,\dots,c_k\}\up$. Moreover, an easy induction shows that
$Z_{i-1}\up\subseteq Z_i\up$ for every $i$ (because both $pre^\forall$
operations preserve inclusion). Once again, because relation $\preceq$
is a well-quasi-order, there has to be $i$ such that
$Z_{i-1}\up=Z_i\up$. Let us write $Z_\infty$ for this $Z_i$.

First, we show that $Z_\infty$ is indeed the complement of $\hGg$.

\begin{lem}\label{lem:zbiory_Z_sa_ok}
  There is a good computation from a configuration $c$ iff
  $c\not\in Z_\infty\up$.
\end{lem}

\proof
  ($\Rightarrow$) We show by induction that $c\not\in Z_i$ for
  $i=-1,0,1\dots$. For $i=-1$ it is obvious.  Assume for contradiction that
  there exists a good computation from $c$, but $c\in
  Z_i\up$. Then there exists $c'\preceq c$ with $c'\in Z_i$. From
  Lemma \ref{lem:zmniejszanie} we know that a good infinite
  computation exists also from $c'$. This computation may first read
  some letters from $\Sigma$, but finally it has to read a letter
  $(\mathtt{delay},\cdot)$, that results in a configuration
  $c_2$. Definition of $Z_i$ tells us that $c_2\in Z_{i-1}\up$. But
  from $c_2$ there is also a good infinite computation, a contradiction.

  ($\Leftarrow$) Assume that every computation (finite or
  infinite) from $c$ reads at most $k$ letters
  $(\mathtt{delay},\cdot)$. An easy induction on $k$ shows that $c\in
  Z_k$.
\qed

To compute $Z_\infty$ it is enough to show how to
compute $Z_i\up$ from $Z_{i-1}\up$. This is the most difficult part of
the proof that will occupy the rest of the subsection. Once this is done
we will calculate all the sets $Z_i\up$, starting with
$Z_{-1}=\emptyset$ and ending when $Z_{i-1}\up=Z_i\up$.

The main idea in calculating
$pre_{\Sigma^*}^\forall(pre^\forall_{\mathtt{delay}}(X))$ is that the
length of its generators may be bounded by some function in the length
of generators of $X$. This is expressed by the following
lemma.

\begin{lem}\label{lemma:bound-pre}
  Given an upward closed set $X$ we can compute a constant $D(X)$
  (which depends also on our fixed automaton $\Aa$) such that the size
  of every minimal element of
  $pre_{\Sigma^*}^\forall(pre^\forall_{\mathtt{delay}}(X))$ is bounded
  by $D(X)$
\end{lem}

Once we know the bound on the size of generators, we can try all
potential candidates. The following lemma shows that it is possible.

\begin{lem}\label{lemma:membership}
  For every upper-closed set $X$, the membership in
  $pre_{\Sigma^*}^\forall(pre^\forall_\mathtt{delay}(X))$ is decidable.
\end{lem}

Together Lemmas~\ref{lemma:bound-pre} and~\ref{lemma:membership}
allow us to compute the sequence $Z_0,Z_1,\dots,Z_\infty$ and hence
also $\hGg$. 

To finish the proof of the theorem, it remains to give proofs of the
two lemmas.  The first is substantially more complicated, and will
occupy most of the space, while the second we will get as a rather
simple corollary. In the first proof, we will calculate separately bounds for
$pre^\forall_\mathtt{delay}(X)$ and for
$pre^\forall_\mathtt{\Sigma^*}(X)$. In the sequel we will need to use
some special representation for sets of configurations.
\begin{defi}
 A \emph{compressed configuration} has a  form
$$\ccc=(\lambda_1\dots\l_l,f,\l_\infty),$$
where $\l_i\in\Lambda_I$, $\lambda_\infty\in\Lambda_\infty$ and
$f:\Lambda_I\to\mathcal{P}(\Lambda_I)$ (values of $f$ are subsets of
$\Lambda_I$).
\end{defi}

On compressed configurations we introduce an expansion operation
pa\-ra\-me\-tri\-zed by words from $\Lambda_I^*$.

\begin{defi}
  A compressed configuration $\ccc=(\lambda_1\dots\l_l,f,\l_\infty)$
  may be expanded in a context of some word
  $\l_1^0\dots\l_k^0\in\Lambda^*_I$, giving as a result the set of configurations
  $(\l_1\dots\l_l\l'_{l+1}\dots\l'_{l+k},\l_\infty)$ such that
  $\l'_{l+i}\in f(\l_i^0)$ for $1\leq i\leq k$.
  We will use $\Exp{\ccc}{\l_1^0\dots\l_k^0}$ to denote the set of
  obtained configurations. Similarly, if $\hCc$ is a set of compressed
  configurations we write $\sExp{\hCc}{\l_1^0\dots\l_k^0}$ for
  $\bigcup\set{\Exp{\ccc}{\l_1^0\dots\l_k^0}: \ccc\in\hCc}$.
\end{defi}

Observe that the value $f(\l)$ for $\l$ not appearing in
$\l_1^0\dots\l_k^0$ does not matter; moreover if some $f(\l_i^0)=\emptyset$
then the result of expanding is the empty set.

We use compressed configurations, because the set of successors of a
configuration may be described by a bounded number of compressed
configurations. This is not true for ordinary configurations due to
nondeterminism. For example, when there is more than one choice of
a transition on action $a$ form a letter $\l$ then every occurrence of
$\l$ in a configuration may make a choice independently, so the number
of successor configurations grows with the number of occurrences of $\l$ in a
configuration. 

Let us see how to calculate $pre^\forall_{\mathtt{delay}}(X)$.  Some
care is needed as this set is not upward closed with respect to the
$\preceq$ relation. This is because the a $(\delay,\cdot)$ action treats the
one before the last element of a configuration in a special way. So if
something is inserted after $\l_k$ in $(\l_1\dots\l_k,\l_\infty)$ then
the $\mathtt{delay}$ operation uses this inserted element instead of
$\l_k$. As a side remark let us mention that using the upward closure
of $pre^\forall_{\mathtt{delay}}(Z_{i-1}\up)$ in the definition of
$Z_i$ would be incorrect (Lemma~\ref{lem:zbiory_Z_sa_ok} would not be
true).

To remedy this problem we use a refined relation
$\preceq_r$\label{def:r-fleq}. Given two configurations
$c'=(\l'_1\dots\l'_{k'},\l'_\infty)$ and $c=(\l_1\dots\l_k,\l_\infty)$
we set
\begin{equation*}
    c' \preceq_r c
  \quad\text{iff}\quad 
      k'>0,\ 
      c' \preceq c
    \text{ and }
      \l'_{k'}\subseteq\l_k
\end{equation*}

Note that the set $pre^\forall_{\mathtt{delay}}(X)$\label{rem:precr}
is upward closed with respect to relation $\preceq_r$, when $X$ is
upward closed with respect to $\preceq$. This is because if
$c'_1\preceq_r c_1$ and $c_1\act{(\mathtt{delay},\cdot)}c_2$ then also
$c'_1\act{(\mathtt{delay},\cdot)}c'_2$ with some $c'_2\preceq
c_2$. Hence, if $c_1\not \in pre^\forall_{\mathtt{delay}}(X) $ then
$c_1'\not\in pre^\forall_{\mathtt{delay}}(X)$.

The following lemma tells us that successors of a configuration may be described
using compressed configurations and that there are not too many of them.

\begin{lem}\label{lemma:coveri}
  For every configuration $c_0=(\l_1\dots\l_k,\l_\infty)$, $k>0$ there
  exists a finite set of compressed configurations $\hCc(\l_k,\l_\infty)$
  (depending only on $\l_k$ and $\l_\infty$) 
  such that:
  \begin{iteMize}{$\bullet$}
  \item if $c_0\act{(\mathtt{delay},\cdot)}c$ then $c\in
    \sExp{\hCc(\l_k,\l_\infty)}{\l_1\dots\l_{k-1}}$;
  \item if $c\in \sExp{\hCc(\l_k,\l_\infty)}{\l_1\dots\l_{k-1}}$ then
    $c_0\act{(\mathtt{delay},\cdot)}c'$ for some $c'\preceq c$.
  \end{iteMize}
\end{lem}

\proof
  The transition on $(\mathtt{delay},\epsilon)$ is deterministic.
  If $c_0\act{(\mathtt{delay},\epsilon)}c'$ then we either have
  $c'=(\l_k'\l_1\dots\l_{k-1},\l_\infty')$ or
  $c'=(\l_1\dots\l_{k-1},\l_\infty')$ depending on $\l_k$. In the
  first case we add $\ccc=(\l_k',\mathbf{sgl},\l_\infty')$ to
  $\hCc(\l_k,\l_\infty)$, in the second case
  $\ccc=(\epsilon,\mathbf{sgl},\l_\infty')$, where
  $\mathbf{sgl}(\l)=\set{\l}$.  In both cases
  $\Exp{\ccc}{\l_1\dots\l_{k-1}}=\{c'\}$.

  Now consider transitions reading $(\mathtt{delay},a)$. A result of
  this transition is not unique and depends on the choice of a
  transition for each element of the configuration. We fix a set
  $\mathcal{T}$ of transitions $\l\act{a}_\Aa(\l',\gamma')$;
  intuitively these are allowed transitions from
  $\l_1,\dots,\l_{k-1}$.  We also fix transitions
  $\l_k^1\act{a}_\Aa(\l_k',\gamma_k')$ and
  $\l_\infty\act{a}_\Aa(\l_\infty',\gamma_\infty')$ (where $\l_k^1$ is
  $\l_k$ with increased regions as in Lemma~\ref{lem:tranz_dla_delay_lit}). This choice
  of transitions gives us a compressed configuration
  $\ccc=(\gamma,f,\l_\infty'')$, where
\begin{align*}
  \gamma=&\
  \gamma_k'\cup\gamma'_\infty\cup\{(q,\Ii_{d+1}):(q,\{d\})\in\l_k',d<d_{max}\}\\
  &\cup\bigcup
  \{\gamma':(\l\act{a}_\Aa(\l',\gamma'))\in\mathcal{T},
  \l\in\Lambda_I\},\\
  f(\l)=&\bigcup\{\l':(\l\act{a}_\Aa(\l',\gamma'))\in\mathcal{T}\},\\
\l''_\infty=&\l'_\infty\cup\set{(q,\Ii_\infty) : (q,\set{d_{max}})\in\l_k'}.
\end{align*}
We add $\ccc$ into $\hCc(\l_k,\l_\infty)$.

We now show that the constructed $\hCc(\l_k,\l_\infty)$ has the
required properties.  Consider a successor $c$ of $c_0$ that is
reached using the transitions we have fixed. In particular, we require
that each transition from $\Tt$ is used at least once. Take $\ccc$ as
calculated above. Directly from the definition we get $c\in
\Exp{\ccc}{\l_1\dots\l_{k-1}}$. As the choice of transitions was
arbitrary, this gives the first statement of the lemma.

Now consider $c=(\gamma\l_1'\dots\l_{k-1}',\l''_\infty)\in
\Exp{\ccc}{\l_1\dots\l_{k-1}}$ where $\ccc=(\gamma,f,\l''_\infty)$ is
obtained by a choice of some $\Tt$ and some transitions from $\l^1_k$
and $\l_\infty$. For every $i$ let us choose some transition
$\l_i\act{a}_\Aa(\l'_i,\gamma'_i)$ from $\mathcal{T}$ (there is at
least one such transition in $\mathcal{T}$ because $\l'_i\in
f(\l_i)$).  Take $c'=(\gamma'\l'_1\dots\l'_{k-1},\l''_\infty)$ where
$$\gamma'=
\gamma_k'\cup\gamma'_\infty\cup\{(q,\Ii_{d+1}):(q,\{d\})\in\l_k',d<d_{max}\}
\cup\bigcup_{1\leq i\leq k-1}\gamma'_i$$ Then $\gamma'\incl \gamma$ so
$c'\preceq c$. It is easy to check that there is a transition
$c_0\act{(\mathtt{delay},a)}c'$.
\qed

We need to find all minimal elements of
$pre^\forall_{\mathtt{delay}}(Z_{i-1}\up)$.  The following lemma
will allow us to get a bound on their size.

\begin{lem}\label{lem:ogran}
  For a given $\hCc$ and a set $X$ upward closed with respect to the $\preceq_r$
  relation there exists a constant $B(X,\hCc)$ (and we may
  compute it) such that if for some $\l_1^0\dots\l_k^0$
$$\sExp{\hCc}{\l_1^0\dots\l_k^0}\subseteq X$$ 
then there exist $1\leq i_1<\dots<i_m\leq k$, $m<B(X, \hCc)$ with
$$\sExp{\hCc}{\l_{i_1}^0\dots\l_{i_m}^0}\subseteq X.$$
\end{lem}

\proof
  First suppose that $\hCc$ is a singleton $\set{\ccc}$; where
  $\ccc=(\l_1\dots\l_l,f,\lambda_\infty)$. We describe a construction
  of a finite
  automaton $\Aa^X_{\ccc}$ accepting the language
  \begin{equation*}
    L^X_{\ccc}=\set{\l_1'\dots\l_k' :
      \Exp{\ccc}{\l_1'\dots\l_k'}\subseteq X}.
  \end{equation*}
  Recall that $X$ is an upward closed set with respect to $\preceq_r$
  relation. This implies that $L^X_{\ccc}$ is upward closed with
  respect to the standard subsequence relation $\sqleq$. It is easy to
  check that for every letter $\l\in\Lambda_I$, if $L\incl
  \Lambda_I^*$ is $\sqleq$-upward closed then the quotient $L/\l$ is
  also $\sqleq$-upward closed. Moreover $L\incl L/\l$, as if $w\in L$
  then $aw\in L$ that implies $w\in L/\l$. Because $\sqleq$ is a well-quasi-order, this last property implies that the set of all possible
  quotients of $L^X_{\ccc}$, i.e. the languages $L^X_{\ccc}/w$ for
  $w\in \Lambda^*_I$, is finite. These quotients are the states of
  $\Aa^X_{\ccc}$ we were looking for. Indeed $\Aa^X_{\ccc}$ is the
  minimal deterministic automaton for $L^X_{\ccc}$. Take $B(X,\set{\ccc})$
  to be the size of the automaton. From the pumping lemma it follows
  that if the word $\l_1^0\dots\l_k^0$ is accepted by $\Aa^X_{\ccc}$
  then there is a subsequence of length $\leq B(X,\set{\ccc})$ accepted by
  $\Aa^X_{\ccc}$.

  Now consider a general situation. For every $\ccc\in\hCc$ from above
  we have some subsequence $\l^0_{i_1}\dots \l^0_{i_m}$ of length
  $m\leq B(X,\{\ccc\})$, such that $\Exp{\ccc}{\l^0_{i_1}\dots
    \l^0_{i_m}}\subseteq X$. We take all the elements from all these
  subsequences, getting a subsequence of length $\leq
B(X,\hCc):=\sum_{\ccc\in\hCc}B(X,\{\ccc\})$
  such that all the inclusions hold.
\qed

The above two lemmas allow to compute a bound on the size of minimal
elements in $pre^\forall_{\mathtt{delay}}(Z_{i-1}\up)$.

\begin{lem}\label{lemma:Mdelay}
  There is an algorithm that given $X\up$ computes a constant
  $M_{delay}(X\up)$ such that the size of every minimal element of
  $pre^\forall_{\mathtt{delay}}(X\up)$ is bounded by $M_{delay}(X\up)$.
\end{lem}
\proof
  There are only finitely many different $\hCc(\l_k,\l_\infty)$ as
  constructed in Lemma~\ref{lemma:coveri}. Let $M_{delay}$ be
  the maximal possible value of $B(X\up,\hCc(\l_k,\l_\infty))$.

  Suppose $c_0=(\l_1^0\dots\l_k^0,\l^0_\infty)$ is a minimal element
  of $pre^\forall_{\mathtt{delay}}(X\up)$. Take the set
  $\hCc(\l^0_k,\l^0_\infty)$ as given by Lemma~\ref{lemma:coveri}. We
  have that
  $\sExp{\hCc(\l^0_k,\l^0_\infty)}{\l_1^0\dots\l_{k-1}^0}\subseteq
  X\up$ by the second statement of this lemma. From
  Lemma~\ref{lem:ogran} we get a subsequence $\l'_1\dots\l'_l$ of
  $\l^0_1\dots\l^0_{k-1}$ whose length is bounded by
  $B(X\up,\hCc(\l_k^0,\l_\infty^0))\leq M_{delay}$ and such that
  $\sExp{\hCc(\l^0_k,\l^0_\infty)}{\l'_1\dots\l'_l}\subseteq
  X\up$. By the first statement of Lemma~\ref{lemma:coveri} we get that
  $(\l'_1\dots\l'_l\l_k^0,\l_\infty^0)\in
  pre^\forall_{\mathtt{delay}}(X\up)$. By the minimality of
  $c_0$, we get that $c_0=(\l'_1\dots\l'_l\l_k^0,\l_\infty^0$), so its
  length is bounded by $M_{delay}+2$.
\qed

Now we describe how to calculate $pre^\forall_{\Sigma^*}(Y)\up$ for any set
$Y$ upward closed with respect to $\preceq_r$ relation. The first lemma
says that we may represent successors using compressed configurations.

\begin{lem}\label{lem:nast}
  For every compressed configuration $\ccc_0$ there is a set of
  compressed configurations $\hCc(\ccc_0)$ (and we may compute it) such
  that for every $\l^0_1\dots\l^0_k$
\begin{iteMize}{$\bullet$}
\item if $c_0\in \Exp{\ccc_0}{\l_1^0\dots\l_k^0}$ and $c_0\act{a}c$
  for some $a\in\Sigma$, then $c\in
  \sExp{\hCc(\ccc_0)}{\l^0_1\dots\l_k^0}$;
\item if $c\in \sExp{\hCc(\ccc_0)}{\l^0_1\dots\l^0_k}$, then
$c_0\act{a}c'$ for some $c'\preceq_r c$, $a\in\Sigma$ and some $c_0\in
\Exp{\ccc_0}{\l^0_1\dots\l^0_k}$.
\end{iteMize}
\end{lem}

\proof
Let $\ccc_0=(\l_1\dots\l_l,f,\l_\infty)$. Fix a letter $a\in \Sigma$.
 We fix a set $\Tt$ of transitions $\l\act{a}_\Aa(\l',\gamma')$; intuitively
these are allowed transitions from $\l\in f(\l^0_i)$. 
We also fix transitions $\l_i\act{a}_\Aa(\l'_i,\gamma'_i)$ for $i=1,\dots,l,\infty$.
This choice of transitions gives us a
compressed configuration
$\ccc=(\gamma\l'_1\dots\l'_l,f',\l'_\infty)$, where
\begin{align*}
  \gamma=&\bigcup_{i=1\dots,l,\infty}\gamma'_i\cup\bigcup
    \{\gamma':(\l\act{a}_\Aa(\l',\gamma'))\in\Tt,\l\in\Lambda_I\},
  \\
    f'(\l^0)=&\{\l':(\l\act{a}_\Aa(\l',\gamma'))\in\Tt,\l\in f(\l^0)\}.
\end{align*}
We add $\ccc$ into $\hCc(\ccc_0)$.

For the first statement of the lemma, take
$c_0\in\Exp{\ccc_0}{\l_1^0\dots\l_k^0}$ and consider any successor $c$ of
$c_0$ that is reached using the transitions we have fixed. In particular we
require that each transition from $\Tt$ is used at least once. 
Take $\ccc$ as calculated above. Then directly from the
definition we get $c\in \Exp{\ccc}{\lambda^0_1\dots\l_k^0}$. As the
choice of transitions was arbitrary this gives the first statement of the
lemma.

Now consider some $\ccc\in \hCc(\ccc_0)$. It is of the form
$(\gamma\l'_1\dots\l'_l,f',\l'_\infty)$. According to the above, it was
constructed from $\ccc_0$ using some transitions
$\l_i\act{a}_\Aa(\l'_i,\gamma'_i)$ for $i=1,\dots,l,\infty$ and some
set of transitions $\Tt$. Take $c\in
\Exp{\ccc}{\l_1^0\dots\l_k^0}$. We have that $c$ is of the form
$(\gamma'\l_1'\dots\l'_l\l'_{l+1}\dots\l'_{l+k},\l'_\infty)$ where
$\l'_1\dots\l'_l$ are as in $\ccc$ and for $i=1,\dots,k$ we can
choose from $\Tt$ transitions
$\l_{l+i}\act{a}_\Aa(\l'_{l+i},\gamma'_{l+i})$ such that $\l_{l+i}\in
f(\l^0_i)$. Take
$c_0=(\l_1\dots\l_l\l_{l+1},\dots,\l_{l+k},\l_\infty)$, i.e. a
configuration whose components are predecessors of transitions we have
selected. We have $c_0\in \Exp{\ccc_0}{\l_1^0\dots\l_k^0}$ by the
definition of expansion. Let
$c'=(\gamma'\l_1'\dots\l'_{l+k},\l'_\infty)$ with
$\gamma'=\bigcup_{i=1,\dots,l+k,\infty}\gamma'_i$. Observe that
$\gamma'$ may be a proper subset of $\gamma$ if not all transitions
from $\Tt$ have been used. Then $c'\preceq_r c$ and there is a
transition $c_0\act{a}c'$.

\qed

The following lemma says that we may list a big enough portion of all
configurations reachable from some $c_0$ (similarly like in step two
of the decision procedure, Lemma~\ref{lemma:reach-tree}) and moreover
  that size of this portion is bounded by a constant.

\begin{lem}\label{lemma:coverii}
  For every $\l_\infty\in\Lambda_\infty$ we can construct a set
  $\hCc_{\Sigma^*}(\l_\infty)$ such that for every
  $\l_1\dots\l_k\in\Lambda^*_I$
  \begin{iteMize}{$\bullet$}
  \item if $(\l_1\dots\l_k,\l_\infty)\actS c$ for some $c$ then there
    is  $c'\preceq_r c$ such that\\
    $c'\in\sExp{\hCc_{\Sigma^*}(\l_\infty)}{\l_1\dots\l_k}$;
  \item if $c\in\sExp{\hCc_{\Sigma^*}(\l_\infty)}{\l_1\dots\l_k}$ then
    there is $c'\preceq_r c$ with $(\l_1\dots\l_k,\l_\infty)\actS c'$. 
  \end{iteMize}
\end{lem}

\proof
  Take the compressed configuration
  $\ccc_0=(\epsilon,\mathbf{sgl},\l_\infty)$, where, as before,
  $\mathbf{sgl}(\l)=\set{\l}$. We define a set $\hCc$ of compressed
  configurations as a closure of $\set{\ccc_0}$ on the operation
  defined in Lemma~\ref{lem:nast}. This set may be infinite but we do
  not worry about it for the moment. We show first that it satisfies
  the requirements of the lemma.

  Take some $\l_1\dots\l_k\in\Lambda^*_I$ and $c$ such that
  $(\l_1\dots\l_k,\l_\infty)\actS c$. We need to show that we can find an
  extended configuration $\ccc\in \hCc$ such that
  $c\in\Exp{\ccc}{\l_1\dots\l_k}$. The proof is by easy induction on
  the number of transitions. For the base step we have
  $(\l_1\dots\l_k,\l_\infty)\in \Exp{\ccc_0}{\l_1\dots\l_k}$, and the
  induction step is given by the first statement of
  Lemma~\ref{lem:nast}.

  Now, suppose that $\ccc\in \hCc$ and $c\in
  \Exp{\ccc}{\l_1\dots\l_k}$. An induction using the second statement
  of Lemma~\ref{lem:nast} shows that there is $c'\preceq_r c$ such that
  $(\l_1\dots\l_k,\l_\infty)\stackrel{\Sigma^*}{\twoheadrightarrow}
  c'$.

  In order to reduce $\hCc$ to a finite set we once again use
  well-quasi-orders. We define a relation $\sqsubseteq$ on compressed
  configurations:
  \begin{eqnarray*}
    \lefteqn{(\l_1'\dots\l'_{l'},f',\l'_\infty)\sqsubseteq
      (\l_1\dots\l_{l},f,\l_\infty)
      \quad\Longleftrightarrow}\\&&
    (\l_1'\dots\l'_{l'},\l'_\infty)\preceq_r(\l_1\dots\l_{l},\l_\infty)
    \textrm{ and } f=f'.
\end{eqnarray*}
This relation is a well-quasi-order. We take $\hCc_{\Sigma^*}(\l_\infty)$
to be the set of minimal elements in this quasi-order. 
It is clear that $\sExp{\hCc_{\Sigma^*}(\l_\infty)}{\l_1\dots\l_k}
\subseteq \sExp{\hCc}{\l_1\dots\l_k}$ for arbitrary $\l_1\dots\l_k$. So, by
the above observations the second property of the lemma holds. For the
first property observe that  whenever
$\ccc'\sqsubseteq \ccc$ and $c\in \Exp{\ccc}{\l_1\dots\l_k}$ then
there is $c'\in \Exp{\ccc'}{\l_1\dots\l_k}$ with $c'\preceq_r c$.  
\qed

\begin{lem}\label{lemma:Msigma}
  There is an algorithm that given a set $Y$ upward closed with respect to the
  $\preceq_r$ relation computes a constant
  $M_{\S^*}(Y)$ such that the size of every minimal element of
  $pre^\forall_{\Sigma^*}(Y)$ is bounded by $M_{\S^*}(Y)$.
\end{lem}
\proof There are only finitely many different $\hCc(\l_\infty)$
constructed in the above lemma. Let $M_{\S^*}$ be the maximal possible
value of $B(Y,\hCc(\l_\infty))$ (cf. Lemma~\ref{lem:ogran})

  Suppose $c_0=(\l_1^0\dots\l_k^0,\l^0_\infty)$ is a minimal element
  of $pre_{\Sigma^*}^\forall(Y)$. Take the set $\hCc(\l^0_\infty)$ as
  given by Lemma~\ref{lemma:coverii}. We have that
  $\sExp{\hCc_{\S^*}(\l^0_\infty)}{\l_1^0\dots\l_{k}^0}\subseteq Y$ by
  the second statement of this lemma. From Lemma~\ref{lem:ogran} we
  get a subsequence $\l'_1\dots\l'_l$ of $\l_1^0\dots\l_k^0$ whose length is
  bounded by $B(X,\hCc(\l^0_\infty))\leq M_{\S^*}$ and such that
  $\sExp{\hCc_{\S^*}(\l^0_\infty)}{\l'_1\dots\l'_l}\subseteq Y$. By the
  first statement of Lemma~\ref{lemma:coverii} we get that
  $(\l'_1\dots\l'_l,\l_\infty^0)\in
  pre_{\Sigma^*}^\forall(Y)$. By the minimality of $c_0$, we have
  that $c_0=(\l'_1\dots\l'_l,\l_\infty^0)$, so its length is bounded by
  $M_{\S^*}+1$.
\qed

The last step before proving Lemmas~\ref{lemma:bound-pre}
and~\ref{lemma:membership} consists of two simple observations.

\begin{lem}\label{lemma:delay-membership}
For every set $X$ upward closed with respect to $\preceq$ relation, the
membership in $Y=pre^\forall_{\ddelay}(X)$ is decidable. Moreover $Y$
is a $\preceq_r$-upward closed set.
\end{lem}
\proof
  The first part of the lemma is obvious, it suffices to test all
  possible transitions that are explicitly characterized in
  Lemmas~\ref{lem:tranz_dla_delay} and~\ref{lem:tranz_dla_delay_lit}.
  The second part follows from the property that we have already
  noticed before (page~\pageref{rem:precr}): if $c'_1\preceq_r c_1$
  and $c_1\act{(\mathtt{delay},\cdot)}c_2$ then also
  $c'_1\act{(\mathtt{delay},\cdot)}c'_2$ with some $c'_2\preceq c_2$.
\qed

\begin{lem}\label{lemma:sigma-membership}
For every set $Y$ upward closed with respect of $\preceq_r$ relation, the
membership in $pre^\forall_{\Sigma^*}(Y)$ is decidable.
\end{lem}
\proof
  Given a configuration $c$ we need to decide if $c\in
  pre_{\Sigma^*}^\forall(Y)$. We apply successively $\act{a}$
  transitions to $c$ constructing a part of the reachability tree. We
  stop the development in a node if it has an ancestor smaller with
  respect to $\preceq_r$-relation. As $\preceq_r$ is a
  well-quasi-order, and the branching at each node is finite, we get a
  finite tree $t$.

  It remains to argue that this construction is correct. If in the
  above process we find a configuration that is not in $Y$ then
  clearly $c$ is not in $pre_{\Sigma^*}^\forall(Y)$. For the other
  direction, assume conversely that there is $c'\notin Y$ with
  $c\mactS c'$. Choose  $c'\notin Y$ so that the length of a
  derivation $c\mactS c'$ is the smallest possible. We show that
  $c'\in t$. Recall that Lemma~\ref{lem:tranz_dla_lit} characterizes
  transitions on letters. Directly from this characterization we
  obtain that if $c'_1\preceq_r c_1$ and $c_1\act{a} c_2$ then also
  $c'_1\act{a} c'_2$ with some $c'_2\preceq_r c_2$. Using this fact,
  we get that if $c'$ is not in $t$ then there is $d'\preceq c'$
  such that the derivation $c\mactS d'$ is shorter than $c\mactS
  c'$. This is impossible by the choice of $c'$.
\qed

\noindent{\upshape\bfseries Proof} (of Lemma~\ref{lemma:bound-pre})\\
Take an upward closed set $X$. By Lemma~\ref{lemma:Mdelay} we can
compute a constant $M_\ddelay$ that bounds the size of minimal
elements in $Y=pre^\forall_\ddelay(X)$. Using
Lemma~\ref{lemma:delay-membership} we can find the minimal elements of
$Y$ by enumerating all configurations of size bounded by
$M_\ddelay$. Observe that $Y$ is $\preceq_r$ upward closed.

Once we have computed $Y$, Lemma~\ref{lemma:Msigma} gives us a
constant $M_{\S^*}(Y)$ bounding the size of minimal elements in
$pre^\forall_{\Sigma^*}(Y)=pre^\forall_{\Sigma^*}(pre^\forall_\ddelay(X))$.

\koniec\medskip

\noindent{\upshape\bfseries Proof} (of Lemma~\ref{lemma:membership})\\
We first compute the set $Y=pre^\forall_\ddelay(X)$ as described
above. We can then use Lemma~\ref{lemma:sigma-membership} to test for
the membership in
$pre^\forall_{\Sigma^*}(Y)=pre^\forall_{\Sigma^*}(pre^\forall_\ddelay(X))$.

\koniec\medskip



\section{Constrained TPTL}\label{sec:logic}

In this section we present a fragment of TPTL (timed propositional
temporal logic) that can be translated to automata whose emptiness
problem is decidable by Theorem~\ref{thm:decidable}.  We compare this
fragment with other known logics for real time. We will be rather
brief in presentations of different formalisms, and refer the reader
to recent surveys~\cite{Bouyer-M4M5,OW08}.

TPTL\cite{AH94} is a timed extension of linear time temporal logic
that allows to explicitly set and compare clock variables. We will
consider the logic with only one clock variable that we denote
TPTL$^1$. The syntax of the logic is:
\begin{equation*}
  p\ |\ x.\a\ |\ x\sim c\ |\ \a\land\beta\ |\ \a\lor\beta\ |\ \a\UU\beta\ |\ \a
    \dUU\beta,
\end{equation*}
where $p$ ranges over action letters, $x$ is the unique clock
variable, and $x\sim c$ is a comparison of $x$ with a constant
with $\sim$ being one of $=,\not=,<,\leq,>,\geq$. We do
not have negation in the syntax, but from the semantics it will be
clear that negation is definable.

The logic is evaluated over timed sequences
$w=(a_1,t_1)(a_2,t_2)\dots$ We define a satisfaction relation
$w,i,v\sat\a$ saying that a formula $\a$ is true at a position $i$ of
a timed word $w$ with a valuation $v$ of the unique clock variable:

\begin{tabular}{ll}
       $w,i,v\sat p$&  if $a_i=p$,\\
       $w,i,v\sat x\sim c$& if  $t_i - v \sim c$,\\
       $w,i,v\sat x.\a$&  if $w,i,t_i\sat\a$,\\
       $w,i,v\sat \a\UU\b$& if $\exists_{j>i}\ ( w,j,v\sat\b$ and
    $\forall_{k\in (i,j)}\ w,k,v\sat\a)$,\\
    $w,i,v\sat \a\dUU\b$& if $\forall_{j>i}\ (w,j,v\sat\b$ or
    $\exists_{k\in (i,j)} w,k,v\sat\a)$.\\
\end{tabular}

As usual, ``until'' operators permit us to  introduce ``sometimes'' and ``always''
operators:
\begin{equation*}
  F\a \equiv \ttrue\UU\a,\qquad G\a \equiv \ffalse\dUU\a.
\end{equation*}
For the following it will be interesting to note that the two ``until''
operators are inter-definable once we have ``always'' and ``sometimes''
operators:
\begin{equation*}
  \a\dUU\b\equiv G\b\lor \b\UU\a,\qquad
  \a\UU\b\equiv F\b\land\b\dUU\a.
\end{equation*}

Observe that TPTL$^1$ subsumes metric temporal logic (MTL). For
example: $\a\UU_{(0,j)}\b$ of MTL is equivalent to $x.(\a\UU
((x<j)\land \b))$. We will not present MTL here, but rather
refer the reader to~\cite{BCM05} where it is also shown that the
following TPTL$^1$ formula is not expressible in MTL
(when considered in the pointwise semantics):
\begin{equation}\label{eq:nonMTL}
  x.(F (b\land F(c\land x\leq 2))).
\end{equation}

The satisfiability problem over infinite timed sequences is
undecidable for MTL~\cite{OW05}, hence also for TPTL$^1$. Using our
decidability result for alternating timed automata, we can
nevertheless find a decidable fragment that we call Constrained
TPTL. The definition of this fragment will use an auxiliary notion of
positive TPTL$^1$ formulas. These formulas can be translated into
alternating automata where all states are accepting.  The set of \emph{positive
  formulas} is given by the following grammar:
\begin{equation*}
  p\ |\ x.\f\ |\ x\sim c\ |\ \f\lor\p\ |\
      \f\land\p\ |\ \f\dUU \p\ |\  F ((x\leq c)\land\p) 
\end{equation*}
The set of formulas of \emph{Constrained TPTL}  is:
\begin{equation*}
  p\ |\ x.\a\ |\ x\sim c\ |\ \a\lor\beta\ |\ \a\land\beta\ | \a\UU \beta\ |\ \f \qquad\text{$\f$ positive}.
\end{equation*}
Observe that the formula~\eqref{eq:nonMTL} belongs to the positive
fragment if we add redundant $(x\leq 2)$ after $b$. 

\begin{thm}
  For a given Constrained TPTL formula $\alpha$ it is decidable whether there is a non Zeno timed word that is a model of
   $\alpha$. The complexity of the problem
  cannot be bounded by a primitive recursive function.
\end{thm}
\proof
 It is enough to give a translation from formulas to automata in the
 class from Theorem~\ref{thm:decidable}. The translation is on the syntax of the
 formula.

 We start with the automaton for positive formulas. The set of states of
 an automaton for a formula will consists of all subformulas of the formula.
 A state
 associated to a formula $\a$ will be denoted by $[\a]$. The intended
 semantics is that a timed word $w$ is accepted from $[\a]$ iff
 $w,1,0\sat\a$.

 The transition relation of the automaton is given in the following
 table.
 \begin{align*}
   [p]&\act{p}\top & [x\sim c]&\dact{*}{x\sim c}\top\\
   [\a\lor\b]&\dact{\e}{}[\a]\lor[\b]  &
   [\a\land\b]&\dact{\e}{}[\a]\land[\b]\\
   [x.\a]&\dact{\e}{x:=0} [\a]\\
   [\a\dUU\beta]&\act{*}[\a]\lor([\beta]\land
   [\a\dUU\beta])\\
   [F\beta]&\act{*}[\beta]\lor[F\beta]
 \end{align*}
 The transitions follow directly the semantics of formulas; state
 $\top$ is a special state from which every timed word is accepted. As
 our automaton is alternating, on the right hand side of the
 transition we can write a boolean expression on successor states.  We
 should also explain labels $*$ and $\e$ over transitions. Transition
 $\act{*}{}$ is just a shorthand for transitions on all letters of the
 alphabet. Transitions $\act{\e}$ and $\dact{\e}{x:=0}$ can be seen as
 eager $\e$-transitions of the automaton: they are executed as soon as
 they are enabled. The other way is to consider them as rewrite rules
 where the real transition of the automaton is obtained at the end of
 the rewriting, i.e., reaching a transition on a letter. In this
 interpretation we should not forget to accumulate resets. For
 example, the above rules give
 \begin{equation*}
   [x.(\a\dUU\beta)]\dact{*}{x:=0}[\a]\lor([\beta]\land[\a\dUU\beta])
 \end{equation*}
 as a ``real'' transition of the automaton.

 All the states are accepting. Notice that in the case of positive
 formulas we will have a state $[F\beta]$ only when $\beta$ is of the
 form $(x\leq c)\land \b'$. As we
 consider only non Zeno words, this assures that the language accepted
 from this state is correct even if the state $[F\beta]$ is accepting. 

 For other formulas of Constrained TPTL we first assume that for every
 positive formula we have already an automaton constructed by the
 above procedure. We then use the clauses above and the clause for
 the $\UU$ operator
 \begin{align*}
   [\a\UU\beta]&\act{\*}[\beta]\lor([\a]\land
   [\a\UU\beta])
 \end{align*}
 to construct the part of the automaton corresponding the 
 remaining formulas.  The accepting states are all those corresponding
 to positive formulas. All the other states are rejecting.

 A standard argument based on induction on the size of the formula
 shows that the translation is correct.  For the complexity bound
 announced in the statement of the theorem, it is enough to check that
 the proof of the same complexity bound for alternating timed automata
 over finite words~\cite{LW08} can be translated into Constrained
 TPTL.  \qed

\subsection{Relation with other logics}

Safety MTL~\cite{OWsafety06} can be seen as an MTL fragment of
positive TPTL. Indeed, both formalisms can be translated to automata
with only accepting states, but the automata obtained from MTL
formulas also have the locality property (cf.~\cite{OWsafety06}). This
property ensures that the clock is always reset when changing
state. The example~\eqref{eq:nonMTL} shows that this is not the case
for positive TPTL. The satisfiability problem for both logics is
non-elementary~\cite{OW07}.

Using equivalences mentioned above FlatMTL\cite{BMOW07} with pointwise
non Zeno semantics can be defined as a set of formulas of the grammar:
\begin{equation*}
  p\ |\ \a\lor\beta\ |\ \a\land\beta\ |\ \a\UU_J\beta\ |\ \chi\UU_I\beta
  |\ \chi 
\qquad \text{$J$ bounded and $\chi\in MITL$,}
\end{equation*}
where MITL is a version of MTL in which we do not allow equality constraints~\cite{alurfedhen96}.
The original definition admits more constructs, but they are redundant
in the semantics we consider.

Both FlatMTL and Constrained TPTL use two different sets of formulas. The MTL
part of the later logic would look like
\begin{equation*}
  p\ |\ \a\lor\beta\ |\ \a\land\beta\ |\ \a\UU_I\beta\ |\ \f 
\qquad \text{$\f$ positive}.
\end{equation*}
From this presentation it can be seen that there are at least two
important differences: (i) constrained TPTL does not have restrictions
on the left hand side of ``until'', and (ii) it uses the positive fragment
instead of MITL. We comment on these two aspects below.

Allowing unrestricted ``until'' makes the logic more expressive but also
more difficult algorithmically. For example, to get the non primitive
recursive bound it is enough to use the formulas generated by the
later grammar without the clause for positive formulas. This should be
contrasted with the \EXPSPACE-completeness result for FlatMTL \cite{BMOW07}.

The use of positive fragment instead of MITL is also important. The
two formalisms are very different in expressive power. The crucial
technical property of MITL is that a formula of the form $\a\UU_I\b$
can change its value at most three times in every unit interval. This
is used in the proof of decidability of FlatMTL, as the MITL part can
be described in a ``finitary'' way. The crucial property of the
positive fragment is that it can express only safety properties (and
all such properties). We can remark that by reusing the construction
of~\cite{OW05} we get undecidability of the positive fragment extended
with a formula expressing that some action appears infinitely
often. Theorem~\ref{thm:undecidable_ne} presented in the next section
implies that this is true even if we do not use punctual constraints
in the positive fragment.  In conclusion, we cannot add MITL to the
positive fragment without losing decidability.



\section{Undecidability without testing for equality}

Ouaknine and Worrell~\cite{OW05} have proved undecidability of MTL over
infinite words in the case of pointwise semantics. Their
construction immediately implies  that the decidability result from the
last section is optimal if classes of accepting conditions are
concerned.

\begin{thm}[Ouaknine, Worrell]\label{thm:undecidable}
  It is undecidable whether a given one-clock universal timed
  automaton $\Aa$ with weak $(1,2)$ conditions accepts some non Zeno
  word.
\end{thm}

Recall that weak parity conditions were defined on page \pageref{weak-parity};
weak $(1,2)$ condition means that each accepting run contains only accepting states, or reaches $\top$.
The construction in op. cit. relies on equality constraints. Indeed,
if we do not allow equality constraints in MTL then we get a fragment
called MITL, and the satisfiability problem for MITL over infinite
words is decidable~\cite{alurfedhen96}.

In this section we would like to show that a similar phenomenon is
very particular to MTL and does not occur in the context of
automata. We show that the undecidability result holds even when 
automata are  only allowed to test if the clock is bigger than $1$.

\begin{thm}\label{thm:undecidable_ne}
  It is undecidable if a given one-clock universal timed
  automaton $\Aa$ with weak $(1,2)$ conditions accepts some non Zeno
  word, even when $\Aa$ does not use tests for equality.
\end{thm}

\emph{Remark:} The above theorems stay true if we replace ``non Zeno
word'' by ``any word''. This is because we can restrict the language
of an automaton to non Zeno words: the set of non Zeno words is
accepted by an automaton with  weak $(1,1)$ conditions.

To prove Theorem \ref{thm:undecidable_ne} we encode a problem of
deciding whether there is a run of a counter machine with insertion
errors satisfying a (strong) B\"uchi condition. This section is split in two
parts. In the first we introduce counter machines with insertion
errors, and show undecidability of the problem in question. In the
second we give an encoding of this problem into the emptiness problem
for automata with weak $(1,2)$ conditions.

\subsection*{Machines with insertion errors}
A \emph{$k$-counter machine with insertion errors} $\Mg$ has
configurations $(q,c^1,\dots,c^k)$ consisting of a control state $q\in
Q$ and values of the counters $c^i\in\mathbb{N}$.  There are three
kinds of transitions: $(q: c^i:=c^i+1;\mathtt{goto}\ q')$ or $(q:
\mathtt{if}\ c^i=0\ \mathtt{then\ goto}\ q')$ or $(q: \mathtt{if}\
c^i>0\ \mathtt{then\ }c^i:=c^i-1;\mathtt{goto}\ q')$.  The set of
transitions $\delta$ of $\Mg$ gives rise to a relation between
configurations, describing a single step of $\Mg$. The machine has
insertion errors, which means that before and after every step it may
increase any of its counters by any value. We will denote this by
$(q,c^1,\dots,c^k)\act{}(q',c'^1,\dots,c'^k)$, to say that we may
reach configuration $(q',c'^1,\dots,c'^k)$ from $(q,c^1,\dots,c^k)$
using some transition from $\delta$ and possibly increasing some
counters before and after the transition. The initial configuration of
the machine $\Mg$ is $(q_0,0,\dots,0)$. Together with the machine
there is given some subset of states $Q_{acc}\subseteq Q$. We say that
a run of $\Mg$ \emph{satisfies the B\"uchi condition} if in infinitely
many of its configurations there appears a state from $Q_{acc}$.

\begin{thm}[Ouaknine,
  Worrell~\cite{OW05}]\label{thm:undecidable_gainy}
  It is undecidable whether a given $5$-counter machine with insertion
  errors $\Mg$ has a run satisfying the B\"uchi condition.
\end{thm}

For completeness, we give a short proof of
Theorem~\ref{thm:undecidable_gainy} by reduction to boundedness of a
lossy $4$-counter machine. The principle of \emph{lossy $k$-counter
  machine} is similar to that with insertion errors, with a difference
that before or after every step it may decrease any of its counters by
any value (instead of increasing). We say that a run of such a machine
is bounded, iff there is a common bound for values of all counters in
all configurations throughout the run. We will use the following result.

\begin{thm}[Mayr~\cite{Mayr03}]\label{thm:undecidable_lossy}
  It is undecidable whether every run of a given lossy $4$-counter
  machine $\Ml$ is bounded.
\end{thm}

\proof[Proof of Theorem \ref{thm:undecidable_gainy}]
Coming back to insertion errors, first note that a counter machine
with insertion errors is exactly the same as lossy counter machine
working backward. Let $\Ml$ be a given lossy $4$-counter machine. We
construct a $5$-counter machine $\Mg$ that can simulate in a backward
fashion a computation of $\Ml$ on the first four counters. This
machine is able to go from a configuration
$(q,c^1,c^2,c^3,c^4,c^5)$ to a configuration $(q_0,0,0,0,0,c^5)$ iff
$\Ml$ can go from $(q_0,0,0,0,0)$, that is the initial configuration,
to $(q,c^1,c^2,c^3,c^4)$.  Additionally to the states of $\Ml$, the
machine has some auxiliary states, among them  an
accepting state $q_{acc}$. The machine will start in the state
$q_{acc}$, and this state will be reachable only from a configuration
$(q_0,0,0,0,0,c^5)$. In the state $q_{acc}$, the machine
increases $c^5$ by $1$ and then (in a nondeterministic way) increases
counters $c^1,c^2,c^3,c^4$, so that $c^1+c^2+c^3+c^4\geq c^5$. To do
that it may the move value of $c^5$ simultaneously into $c^1$ and
$c^2$, then move value from $c^2$ back to $c^5$ and finally while
decreasing $c^1$ increase $c^2,c^3,c^4$. After that it chooses a state
of $\Ml$ and starts computing backward (using only the first four
counters). When configuration $(q_0,0,0,0,0,c^5)$ is reached we make
the machine to go to $(q_{acc},0,0,0,0,c^5)$.

  Assume that $\Ml$ has an unbounded computation. We will show that $\Mg$
  has a run visiting $q_{acc}$ infinitely often. Suppose that some
  initial fragment of this run is already constructed and we are in a
  configuration $(q_{acc},0,0,0,0,c^5)$ for some value of $c^5$. As
  $\Ml$ has an unbounded computation, it can reach a configuration
  $(q,c^1,c^2,c^3,c^4)$ with the sum of the counters bigger than
  $c^5+1$. We increase $c^5$ by 1, distribute $c^5$ into other
  counters to get the values $c^1, c^2, c^3, c^4$, we choose the state $q$ and
  then execute the computation of $\Ml$ backwards, starting from
  $(q,c^1, c^2, c^3, c^4)$. When reaching $(q_0,0,0,0,0,c^5+1)$ we go
  to $(q_{acc},0,0,0,0,c^5+1)$ and repeat this process. This gives the
  required infinite computation.

  For the opposite direction, assume that there is a computation of
  $\Mg$ satisfying the B\"uchi condition. Every appearance of
  $q_{acc}$ is followed by some initialization, and by a backward
  computation of $\Ml$, starting in a configuration of size bigger
  than the value of $c^5$ and ending in $(q_0,0,0,0,0)$. However,
  every time this happens the value of $c^5$ increases by at least
  one.  So we get computations of $\Ml$ ending in bigger and bigger
  configurations.  By K\"onig's lemma, there exists also an unbounded
  computation of $\Ml$.
\qed

\subsection*{Encoding machines into alternating automata}
  Now we return to the proof of Theorem \ref{thm:undecidable_ne}, which occupies the rest of this section.  For
  given $5$-counter machine with insertion errors $\Mg$ we will
  construct an alternating one-clock timed automaton $\Aa$ that
  accepts some infinite word iff $\Mg$ has a run satisfying the
  B\"uchi condition. The input alphabet of $\Aa$ will consist of the
  instructions of $\Mg$ and some auxiliary letters whose use will be
  explained later,
\begin{equation*}
\Sigma=\delta\cup\{\mathtt{shc},\mathtt{sh\$},\mathtt{new},\mathtt{init}\}.
\end{equation*}
As states of $\Aa$ we take
\begin{equation*}
Q_+=Q_\Mm\cup\{1,2,3,4,5,\$,q_\infty,q_{init}\}\quad\text{and}
\quad Q_-=\{q_-\}.  
\end{equation*}
States $Q_\Mm\cup\set{1,2,3,4,5}$ will be used to represent
configurations of $\Mg$: the current state and the values of the five
counters. States $q_\infty$ and $q_-$ will encode the condition on
successful runs. State $\$$ is important for technical reasons
explained later. State $q_{init}$ is just the initial state that will
not be reachable from other states.

In our description below we will consider the characterization of
acceptance given by Lemma~\ref{lemma:acceptance}. In this presentation
a run of $\Aa$ is a sequence
\begin{equation*}
  P_1\hact{a_1,t_1}   P_2\hact{a_2,t_2} P_3\dots,  
\end{equation*}
where each $P_i\incl \Pp(Q_\Aa\times\RR^+)$ is a set of pairs $(q,v)$
consisting of a state of $\Aa$ and a valuation of the clock. We call
such a set an \emph{extended configuration} of $\Aa$, or
an \emph{e-configuration} for short. Compared with
Lemma~\ref{lemma:acceptance} we have joined together a transition
letting the time pass with an action transition and write just
$\hact{a,t}$ transitions. In what follows we will use only two
regions: $\Ii_1=[0,1]$ and $\Ii_\infty=(1,\infty)$.

\begin{defi}\label{def:well-formed}
An e-configuration $P$ of $\Aa$ is \emph{well-formed} if:
\begin{iteMize}{$\bullet$}
  \item For every $(q,v)\in P$: if $q\in \set{1,\dots,5,\$}$ then
    $v\in\Ii_1$, and $v\in\Ii_\infty$ otherwise.
  \item For every $v\in\Ii_1$ there is at most one $q$ with
    $(q,v)\in P$.  
  \item In $P$ there is exactly one pair with a state from $Q_\Mm$,
    exactly one pair with the state $q_\infty$, and no pairs with
    $q_{init}$.
  \item Suppose $(q,v)$ is in $P$ where $q\in\set{1,\dots,5}$. Then
    this pair is immediately preceded by some $(\$,v')$ (there is no
    pair $(q'',v'')$ in $P$ with $v'<v''<v$).
  \end{iteMize}
\end{defi}

\noindent Intuitively, a well-formed e-configuration is divided into two parts:
the set of pairs with the clock value in $\Ii_1$ and those in
$\Ii_\infty$. The first part can be seen as representing a word over
$\set{1,\dots,5,\$}$ that is obtained by using the standard order on
clock values. From the conditions above it follows that this word is
of the form $\$^+q_{i_1}\$^+q_{i_2}\dots\$^+q_{i_n}\$^*$; where
$q_{i_k}\in\set{1,\dots,5}$. Such a word represents values of the
counters when the value of the counter $c^j$ is equal to the number of
$j$ in the word. The clock values of pairs in $\Ii_\infty$ will not
matter, so this part can be seen as a multiset of states. In this
multiset there will be exactly one state from $Q_\Mm$ representing the
state of the simulated machine. State $q_\infty$ plus some number of
states $q_-$ will be there to encode a condition on a successful run.

The automaton $\Aa$ will pass also through e-configurations that are not
well-formed, but in its accepting run it will have to repeatedly
return to well-formed e-configurations. 

\begin{exa}\label{example1}
  Consider an  e-configuration
  $$\{\$^{0.1},1^{0.2},\$^{0.3},\$^{0.4},2^{0.6},\$^{0.8},1^{0.9},q^5,
        q_{-}^5,q_\infty^5\}$$
  where for readability we write a pair $(\$,0.1)$ as $\$^{0.1}$; and
  similarly for all other elements of the set.  This e-configuration
  is well-formed 
  and encodes the configuration $(q,2,1,0,0,0)$ of $\Mm^g$.  Observe
  that there
  are infinitely many well-formed e-configurations encoding this
  configuration of $\Mm$.
\end{exa}

Now we describe transitions of the automaton. In order to have an
intuition for reading the rules below it is important to observe that
if the automaton reads a letter $\s$ then all states in its current
e-configuration have to make a transition according to some rule labeled
$\s$. In consequence, if there is a state in the e-configuration that
does not have a rule for $\s$ then the automaton cannot read $\s$.

The automaton starts in the state $q_{init}$ and waits at least one
time unit to start its two copies: one in a state $q_0$ and another in
$q_\infty$ (where $q_0$ is the initial state of $\Mg$),
   $$q_{init},\Ii_\infty\act{\mathtt{init}}q_0\wedge q_\infty.$$
This means that the e-configuration becomes
$\set{(q_0,v),(q_\infty,v)}$ with $v\in \Ii_\infty$.

States $\$$ for clock values $\leq 1$ are preserved by any transition,
\begin{equation*}
  \$,\Ii_1\act{\sigma}\$,\qquad  \forall\sigma\in\Sigma.
\end{equation*}
Similarly states $1,\dots,5$, with the exception that a transition
checking for zero should not be possible if the corresponding counter is
non-zero,
\begin{equation*}
  i,\Ii_1\act{\sigma}i\qquad  \forall i=1,\dots,5\ 
  \forall \sigma\not=(q: \mathtt{if}\ c^i=0\ \mathtt{then\ goto}\
  q').
\end{equation*}
 When the clock value for a pair with $\$$ or $i$ becomes
  greater than $1$, it may be reset,
$$\begin{array}{ll}
  \$,\Ii_\infty\act{\mathtt{sh\$}}(\$,\mathtt{reset}),&\\
  i,\Ii_\infty\act{\mathtt{shc}}\$\wedge(i,\mathtt{reset})
     &\forall i=1,\dots,5,\\
  q,\Ii_\infty\act{\sigma}q
      &q\in Q_\Mm\cup\set{q_\infty,q_-},\ \sigma=\mathtt{sh\$}\textrm{
        or }\sigma=\mathtt{shc}.
\end{array}$$
Note that the transition on $\$$ reads a different letter than that on
$i$. In consequence, if in a e-configuration there are pairs with both
$\$$ and $i$ having clock values in $\Ii_\infty$ then neither $\shd$ nor
$\shc$ are possible. As we will have no more transitions from $(\$,\Ii_\infty)$
this means that the automaton will be blocked in such e-configuration.

Now we consider moves on transitions of the machine $\Mg$. For $\sigma=(q:
\mathtt{if}\ c^i=0\ \mathtt{then\ goto}\ q')$ we just do
  $$q,\Ii_\infty\act{\sigma}q'.$$
Note that, thanks to earlier restriction, the transition is possible only when
there are no $i$ states in the e-configuration. For 
$\sigma=(q: \mathtt{if}\ c^i>0\ \mathtt{then\ }c^i:=c^i-1;\mathtt{goto}\ q')$
we do
\begin{align*}
  q,\Ii_\infty\act{\sigma}&q',\\
  i,\Ii_\infty\act{\sigma}&\top.
\end{align*}
For $\sigma=(q: c^i:=c^i+1;\mathtt{goto}\ q')$ we do
  $$q,\Ii_\infty\act{\sigma}q'\wedge\$\wedge(i,\mathtt{reset}).$$
As the machine should allow insertion errors, we add a transition
  $$q,\Ii_\infty\act{\mathtt{new}}q\wedge\$\wedge(i,\mathtt{reset}).$$

Finally, we have special states $q_\infty$ and $q_-$, that are used to
ensure that states from $Q_{acc}$ appear infinitely often. The state
$q_\infty$ produces repeatedly new $q_-$ states,
  $$q_\infty,\Ii_\infty\act{\sigma}q_\infty\wedge q_-
  \qquad \forall\sigma\in\Sigma.$$ The state $q_-$ is the only one,
  which is in $Q_-$, so in the accepting run every $q_-$ state has to
  disappear after some time. States $q_-$ disappear, when
  there is a transition ending in a state from $Q_{acc}$,
$$\begin{array}{ll}
  q_-,\Ii_\infty\act{\sigma}\top
      &\forall\sigma=(\dots\texttt{goto}\ q'), q'\in Q_{acc},\\
  q_-,\Ii_\infty\act{\sigma}q_- &\textrm{for all other }\sigma.
\end{array}$$

\begin{exa}
  Let us see how a transition $\sigma=(q: \mathtt{if}\ c^2>0\
  \mathtt{then\ }c^2:=c^2-1;\mathtt{goto}\ q_2)$ is simulated from the
  e-configuration in Example \ref{example1}.  One possibility is to
  immediately execute a transition reading $\sigma$.  We get the
  e-configuration
  \begin{equation*}
      \{\$^{0.1},1^{0.2},\$^{0.3},\$^{0.4},2^{0.6},\$^{0.8},1^{0.9},q_2^5,
        q_{-}^5,q_\infty^5\}
  \end{equation*}
  which is well-formed and encodes the configuration $(q_2,2,1,0,0,0)$
  of $\Mm^g$.  The second counter has not been decreased, but this is
  correct as the machine is allowed to do incremental errors.

  If we really want to decrease the second counter, we have to ensure
  that a pair with $2$ is in the $\Ii_\infty$ region. If we let pass,
  say, $0.2$ units of time we get e-configuration
  \begin{equation*}
      \{\$^{0.3},1^{0.4},\$^{0.5},\$^{0.6},2^{0.8},\$^{1},1^{1.1},q^{5.2},
        q_{-}^{5.2},q_\infty^{5.2}\}\ .
  \end{equation*}
  Then we execute a transition $\mathtt{shc}$ and we get
    \begin{equation*}
      \{1^0,\$^{0.3},1^{0.4},\$^{0.5},\$^{0.6},2^{0.8},\$^{1},q^{5.2},
        q_{-}^{5.2},q_\infty^{5.2}\}\ .
  \end{equation*}
  After time $0.1$ we can execute a transition $\mathtt{sh\$}$, getting
  \begin{equation*}
    \{\$^0,1^{0.1},\$^{0.4},1^{0.5},\$^{0.6},\$^{0.7},2^{0.9},q^{5.3},
        q_{-}^{5.3},q_\infty^{5.3}\}\ .
  \end{equation*}
  Then after $0.2$ we execute a transition $\sigma$, getting a
  well-formed e-configuration corresponding to $(q_2,2,0,0,0,0)$:
  \begin{equation*}
    \{\$^{0.2},1^{0.3},\$^{0.6},1^{0.7},\$^{0.8},\$^{0.9},q_2^{5.5},
        q_{-}^{5.5},q_\infty^{5.5}\}\ .
  \end{equation*}

  Now consider the transition $\sigma'=(q_2: c^3:=c^3+1;\mathtt{goto}\
  q_3)$ of $\Mm^g$.  We execute a transition $\sigma'$ from the
  above e-configuration after time $0.1$ (recall that executing two
  transitions at the same time is forbidden), getting
  \begin{equation*}
    \{3^0,\$^{0.3},1^{0.4},\$^{0.7},1^{0.8},\$^{0.9},\$^{1},q_3^{5.6},
        q_{-}^{5.6},q_\infty^{5.6}\}\ .
  \end{equation*}
  After additional time $0.1$ we execute the transition
  $\mathtt{sh\$}$, getting a well-formed e-configuration corresponding
  to $(q_3,2,0,1,0,0)$:
  \begin{equation*}
    \{\$^0,3^{0.1},\$^{0.4},1^{0.5},\$^{0.8},1^{0.9},\$^{1},q_3^{5.7},
        q_{-}^{5.7},q_\infty^{5.7}\}\ .
  \end{equation*}
\end{exa}

\begin{lem}\label{lemma:gainy-to-aa}
  There exists a run of $\Mg$ satisfying the B\"uchi condition iff
  $\Aa$ accepts some infinite word.
\end{lem}

\proof
  Assume that $\Mg$ has a run satisfying the B\"uchi condition. From
  the initial state, $\Aa$ may go to a well-formed e-configuration
  corresponding to the initial configuration of $\Mg$. Then every step
  of $\Mg$ may be simulated by $\Aa$: When $\Mg$ increases some of its
  counters, we may do the same using transitions on letters
  $\mathtt{new}$ and then $\mathtt{sh\$}$. When $\Mg$ executes a
  transition $\sigma=(q: \mathtt{if}\ c^i=0\ \mathtt{then\ goto}\ q')$
  we may do the same in $\Aa$ reading letter $\sigma$. When $\Mg$ does
  $\sigma=(q: c^i:=c^i+1;\mathtt{goto}\ q')$, we do the same reading
  letter $\sigma$ and then $\mathtt{sh\$}$. It is easy to check, that
  after each step the resulting e-configuration remains well-formed.

  The only complicated transition is $\sigma=(q: \mathtt{if}\ c^i>0\
  \mathtt{then\ }c^i:=c^i-1;\mathtt{goto}\ q')$. Suppose that the
  automaton is in a well-formed e-configuration $P$. Let us look at
  the biggest valuation $v\leq 1$ appearing in $P$. By the conditions
  of well-formedness (c.f.~Definition~\ref{def:well-formed}) there is
  exactly one state $q\in Q_+$ such that $(q,v)\in P$. This state can
  be one of $1,\dots,5,\$$.  The automaton lets the time pass so that
  $v$ becomes greater than $1$, but all other valuations from $\Ii_1$
  stay in $\Ii_1$.  If $q=i$ then the automaton does
  $\sigma$. Otherwise, it does $\shd$ or $\shc$ followed by $\shd$
  that has an effect of putting $\$$ or $\$$ followed by $q$ at the
  beginning of the e-configuration. After this we obtain a well-formed
  e-configuration where the one but the maximal valuation before
  became the maximal one. These operations are repeated until
  $q=i$. We are sure that this process ends, as there is a state $i$
  in $P$.

  To ensure that the obtained word is nonZeno, we have to wait some
  time after every transition of $\Mg$, doing $\mathtt{shc}$ and
  $\mathtt{sh\$}$ if necessary. Observe that every state $q_-$ would
  disappear when in the computation of $\Mg$ there is a
  transition ending in a state from $Q_{acc}$. As this computation
  satisfies the B\"uchi condition, this will happen infinitely often.

 For the other direction, consider an accepting run of $\Aa$ on some
 word.  In the first step, $\Aa$ has to reach a well-formed e-configuration
 corresponding to the initial configuration of $\Mg$. Let us see what may
 happen from any well-formed e-configuration. Suppose that time passes
 and the clock value for some states $1,\dots,5,\$$ becomes greater than
 $1$. If it happens simultaneously for state $\$$ and some state $i$,
 then from the obtained e-configuration there will be no more
 transitions. If it happens only for state $\$$, then the only
 possible transition is the one reading $\mathtt{sh\$}$ after which
 we go back to a well-formed e-configuration corresponding to the same
 configuration of $\Mg$. If it happens just for some state $i$, then
 the automaton can read either $\mathtt{shc}$ or some $(q:
 \mathtt{if}\ c^i>0\ \mathtt{then\ }c^i:=c^i-1;\mathtt{goto}\
 q')$. If it reads $\mathtt{shc}$, then after that it has to read
 $\mathtt{sh\$}$, and we also are back in a well-formed e-configuration
 corresponding to the same configuration of $\Mg$. If it reads
 $\sigma=(q: \mathtt{if}\ c^i>0\ \mathtt{then\
 }c^i:=c^i-1;\mathtt{goto}\ q')$, then we immediately get a
 well-formed e-configuration.

 Transitions reading $\mathtt{shc}$ or $\mathtt{sh\$}$ when no state of
 $1,\dots,5,\$$ has the clock value above $1$ does not change the
 configuration. A transition reading $\mathtt{new}$ has to be followed
 by a transition $\mathtt{sh\$}$ and we get a well-formed e-configuration
 with one of the counters increased. A transition reading $\sigma=(q:
 \mathtt{if}\ c^i=0\ \mathtt{then\ goto}\ q')$ is possible only when
 counter $c^i$ is zero. After a transition reading $\sigma=(q:
 c^i:=c^i+1;\mathtt{goto}\ q')$ there has to be a transition reading
 $\mathtt{sh\$}$ and we get a well-formed e-configuration that
 corresponds to a correct configuration of $\Mg$.  A transition reading
 $\sigma=(q: \mathtt{if}\ c^i>0\ \mathtt{then\
 }c^i:=c^i-1;\mathtt{goto}\ q')$ always gives us a well-formed
 e-configuration. The obtained e-configuration correctly represents the
 result but for the fact that the counter $i$ may not be
 decremented. This is not a problem as we are simulating a machine
 with insertion errors, so we can suppose that the incrementation
 error has occurred immediately after execution of this instruction.

 The above argument gives some computation of $\Mg$ constructed from
 an accepting computation of $\Aa$. So every $q_-$ disappears after
 some time on that computation of $\Aa$. This is only possible when
 reading a letter of the form $(\dots\mathtt{goto}\ q')$ with $q'$ an
 accepting state of $\Mg$. As $q_-$ needs to disappear infinitely
 often, the obtained computation of $\Mg$ is an infinite computation
 satisfying the B\"uchi condition.
\qed

As the choice of counter machine $\Mg$ was arbitrary and the
construction of $\Aa$ from $\Mg$ was effective,
Lemma~\ref{lemma:gainy-to-aa} implies Theorem~\ref{thm:undecidable_ne}.

\section{Conclusions}

This paper presents a study of the emptiness problem for alternating
timed automata. It gives a characterization of decidable cases of this
problem in terms of the complexity of acceptance conditions. The main
result shows that all the classes whose decidability has been left
open are indeed decidable. This result gives new decidability results
for logics for real-time.

Given this characterization, in order to find other, bigger, classes
of alternating timed automata with decidable emptiness problem we need
to look closer at the structure of automata. In this paper one case has been
studied, namely when no punctual constraints are used. This case was
motivated by the phenomenon observed for metric temporal logic: while
the logic is undecidable, it becomes decidable when punctual
constraints are disallowed. The second main result of the paper shows
that in the case of automata such a simple restriction does not
work: one does not  get a bigger decidable class even if one restricts
to extremely simple constraints. This indicates that in order to obtain
larger decidable classes, the structure of resets should be also
examined more closely.



\bibliographystyle{abbrv}
\bibliography{bibliography}

\begin{thebibliography}{10}

\bibitem{AJ98}
P.~Abdulla and B.~Jonsson.
\newblock Veryfying networks of timed processes.
\newblock In {\em Proc. TACAS'98}, volume 1384 of {\em LNCS}, pages 298--312,
  1998.

\bibitem{AJ01}
P.~Abdulla and B.~Jonsson.
\newblock Timed {P}etri nets and {BQO}s.
\newblock In {\em Proc. ICATPN'01}, pages 53--70, 2001.

\bibitem{ADOQW08}
P.~A. Abdulla, J.~Deneux, J.~Ouaknine, K.~Quaas, and J.~Worrell.
\newblock Universality analysis for one-clock timed automata.
\newblock {\em Fundam. Inform.}, 89(4):419--450, 2008.

\bibitem{AOQW07}
P.~A. Abdulla, J.~Ouaknine, K.~Quaas, and J.~Worrell.
\newblock Zone-based universality analysis for single-clock timed automata.
\newblock In {\em FSEN'07}, number 4767 in LNCS, pages 98--112, 2007.

\bibitem{AluDil94}
R.~Alur and D.~Dill.
\newblock A theory of timed automata.
\newblock {\em Theoretical Computer Science}, 126:183--235, 1994.

\bibitem{alurfedhen96}
R.~Alur, T.~Feder, and T.~A. Henzinger.
\newblock The benefits of relaxing punctuality.
\newblock {\em J. ACM}, 43(1):116--146, 1996.

\bibitem{AluFixHen97}
R.~Alur, L.~Fix, and T.~Henzinger.
\newblock Event-clock automata: A determinizable class of timed automata.
\newblock {\em Theoretical Computer Science}, 204, 1997.

\bibitem{AH94}
R.~Alur and T.~A. Henzinger.
\newblock A really temporal logic.
\newblock {\em J. ACM}, 41(1):181--204, 1994.

\bibitem{Bouyer-M4M5}
P.~Bouyer.
\newblock Model-checking timed temporal logics.
\newblock In {\em {W}orkshop on {M}ethods for {M}odalities ({M4M-5})},
  Electronic Notes in Theoretical Computer Science, Cachan, France, 2009.
  Elsevier Science Publishers.
\newblock To appear.

\bibitem{BCM05}
P.~Bouyer, F.~Chevalier, and N.~Markey.
\newblock On the expressiveness of {TPTL} and {MTL}.
\newblock In {\em FSTTCS'05}, volume 3821 of {\em LNCS}, pages 432--443, 2005.

\bibitem{BMOSW08}
P.~Bouyer, N.~Markey, J.~Ouaknine, P.~Schnoebelen, and J.~Worrell.
\newblock On termination for faulty channel machines.
\newblock In {\em STACS'08}, volume 08001 of {\em Dagstuhl Seminar
  Proceedings}, pages 121--132, 2008.

\bibitem{BMOW07}
P.~Bouyer, N.~Markey, J.~Ouaknine, and J.~Worrell.
\newblock The cost of punctuality.
\newblock In {\em LICS'07}, pages 109--120, 2007.

\bibitem{BMOW08}
P.~Bouyer, N.~Markey, J.~Ouaknine, and J.~Worrell.
\newblock On expressiveness and complexity in real-time model checking.
\newblock In {\em ICALP'08}, volume 5126 of {\em LNCS}, pages 124--135, 2008.

\bibitem{HR04}
Y.~Hirshfeld and A.~M. Rabinovich.
\newblock Logics for real time: Decidability and complexity.
\newblock {\em Fundam. Inform.}, 62(1):1--28, 2004.

\bibitem{HJ96}
D.~V. Hung and W.~Ji.
\newblock On the design of hybrid control systems using automata models.
\newblock In {\em FSTTCS'96}, number 1180 in LNCS, pages 156--167, 1996.

\bibitem{LasWal05}
S.~Lasota and I.~Walukiewicz.
\newblock Alternating timed automata.
\newblock In {\em FOSSACS'05}, number 3441 in Lecture Notes in Computer
  Science, pages 250--265, 2005.

\bibitem{LW08}
S.~Lasota and I.~Walukiewicz.
\newblock Alternating timed automata.
\newblock {\em ACM Trans. Comput. Log.}, 9(2), 2008.

\bibitem{Mayr03}
R.~Mayr.
\newblock Undecidable problems in unreliable computations.
\newblock {\em Theoretical Computer Science}, 1-3(297):337--354, 2003.

\bibitem{Mos91}
A.~W. Mostowski.
\newblock Hierarchies of weak automata and week monadic formulas.
\newblock {\em Theoretical Computer Science}, 83:323--335, 1991.

\bibitem{Murlak08}
F.~Murlak.
\newblock Weak index versus borel rank.
\newblock In {\em STACS'08}, Dagstuhl Seminar Proceedings, pages 573--584.
  Dagsr, 2008.

\bibitem{OW04}
J.~Ouaknine and J.~Worrell.
\newblock On the language inclusion problem for timed automata: Closing a
  decidability gap.
\newblock In {\em Proc. LICS'04}, pages 54--63, 2004.

\bibitem{OW05}
J.~Ouaknine and J.~Worrell.
\newblock On the decidability of metric temporal logic.
\newblock In {\em LICS'05}, pages 188--197, 2005.

\bibitem{OW06}
J.~Ouaknine and J.~Worrell.
\newblock On metric temporal logic and faulty {T}uring machines.
\newblock In {\em FoSSaCS}, volume 3921 of {\em LNCS}, pages 217--230, 2006.

\bibitem{OWsafety06}
J.~Ouaknine and J.~Worrell.
\newblock Safety metric temporal logic is fully decidable.
\newblock In {\em TACAS'06}, number 3920 in LNCS, pages 411--425, 2006.

\bibitem{OW07}
J.~Ouaknine and J.~Worrell.
\newblock On the decidability and complexity of metric temporal logic over
  finite words.
\newblock {\em Logical Methods in Computer Science}, 3(1), 2007.

\bibitem{OW08}
J.~Ouaknine and J.~Worrell.
\newblock Some recent results in metric temporal logic.
\newblock In {\em FORMATS'08}, number 5215 in LNCS, pages 1--13, 2008.

\bibitem{VarWop84a}
M.~Y. Vardi and P.Wolper.
\newblock Automata theoretic techniques for modal logics of programs.
\newblock In {\em Sixteenth ACM Symposium on the Theoretical Computer Science},
  1984.

\bibitem{Wag77}
K.~Wagner.
\newblock Eine topologische {C}harakterisierung einiger {K}lassen regul\"{a}rer
  {F}olgenmengen.
\newblock {\em J. Inf. Process. Cybern. EIK}, 13:473--487, 1977.

\bibitem{WS74}
K.~Wagner and L.~Staiger.
\newblock Automatentheoretische und automatenfreie charakterisierungen
  topologischer klassen regularer folgenmengen.
\newblock {\em EIK}, 10:379--392, 1974.

\bibitem{WilkeHab89}
T.~Wilke.
\newblock Classifying discrete temporal properties.
\newblock Habilitation thesis, Kiel, Germany, 1998.

\end{thebibliography}

\end{document}